\documentclass[twocolumn]{aastex62}
\usepackage{natbib}

\usepackage{amsmath}
\usepackage{comment}
\newcommand{\ii}{~\textsc{ii}}
\newcommand{\iii}{~\textsc{iii}}

\newcommand{\samplesize}{264}
\newcommand{\redshiftrange}{0.6 $<$ z $<$ 2.6}
\newcommand{\massrange}{8.5 $<$ $\log$ M$_*$/M$_{\odot}$ $<$ 10.5}

\graphicspath{{./}{figures/}}

\begin{document}

\title{CLEAR: The Gas-Phase Metallicity Gradients of Star-Forming Galaxies at 0.6 $<$ z $<$ 2.6}

\correspondingauthor{Raymond C. Simons}
\email{rsimons@stsci.edu}

\author[0000-0002-6386-7299]{Raymond C. Simons}
\affil{Space Telescope Science Institute, 3700 San Martin Drive,
Baltimore, MD, 21218 US}

\author[0000-0001-7503-8482]{Casey Papovich}
\affiliation{Department of Physics and Astronomy, Texas A\&M University, College
Station, TX, 77843-4242 USA}
\affiliation{George P.\ and Cynthia Woods Mitchell Institute for
 Fundamental Physics and Astronomy, Texas A\&M University, College
 Station, TX, 77843-4242 USA}
 
\author[0000-0003-1665-2073]{Ivelina Momcheva}
\affil{Space Telescope Science Institute, 3700 San Martin Drive,
Baltimore, MD, 21218 US}

\author[0000-0002-1410-0470]{Jonathan R. Trump}
\affil{Department of Physics, University of Connecticut, Storrs, CT 06269, USA}

\author[0000-0003-2680-005X]{Gabriel Brammer}
\affil{Cosmic Dawn Centre, University of Copenhagen, Blegdamsvej 17, 2100 Copenhagen, Denmark}

\author[0000-0001-8489-2349]{Vicente Estrada-Carpenter}
\affiliation{Department of Physics and Astronomy, Texas A\&M University, College
Station, TX, 77843-4242 USA}
\affiliation{George P.\ and Cynthia Woods Mitchell Institute for
Fundamental Physics and Astronomy, Texas A\&M University, College
Station, TX, 77843-4242 USA}

\author[0000-0001-8534-7502]{Bren E. Backhaus}
\affil{Department of Physics, University of Connecticut, Storrs, CT 06269, USA}

\author[0000-0001-7151-009X]{Nikko J. Cleri}
\affil{Department of Physics, University of Connecticut, Storrs, CT 06269, USA}

\author[0000-0001-8519-1130]{Steven L. Finkelstein}
\affil{Department of Astronomy, The University of Texas, Austin, Texas, 78712 USA} 

\author[0000-0002-7831-8751]{Mauro Giavalisco}
\affil{Astronomy Department, University of Massachusetts,
Amherst, MA, 01003 USA} 

\author[0000-0001-7673-2257]{Zhiyuan Ji}
\affil{Department of Astronomy, University of Massachusetts Amherst, 710 N. Pleasant St., Amherst, MA, 01003, USA}

\author[0000-0003-1187-4240]{Intae Jung}
\affil{Department of Physics, The Catholic University of America, Washington, DC 20064, USA}
\affil{Astrophysics Science Division, Goddard Space Flight Center, Greenbelt, MD 20771, USA}

\author[0000-0002-7547-3385]{Jasleen Matharu}
\affiliation{Department of Physics and Astronomy, Texas A\&M University, College
Station, TX, 77843-4242 USA}
\affiliation{George P.\ and Cynthia Woods Mitchell Institute for
Fundamental Physics and Astronomy, Texas A\&M University, College
Station, TX, 77843-4242 USA}

\author[0000-0001-6065-7483]{Benjamin Weiner}
\affil{MMT/Steward Observatory, 933 N. Cherry St., University of Arizona, Tucson,
AZ 85721, USA}

\begin{abstract}
We report on the gas-phase metallicity gradients of a sample of \samplesize{} star-forming galaxies at \redshiftrange{}, measured through deep near-infrared {\emph{Hubble Space Telescope}} slitless spectroscopy. The observations include 12-orbit depth {\emph{Hubble}}/WFC3 G102 grism spectra taken as a part of the CANDELS Ly$\alpha$ Emission at Reionization (CLEAR) survey, and archival WFC3 G102+G141 grism spectra overlapping the CLEAR footprint. The majority of galaxies (84\%) in this sample are consistent with a zero or slightly positive metallicity gradient across the full mass range probed (\massrange{}). We measure the intrinsic population scatter of the metallicity gradients, and show that it increases with decreasing stellar mass---consistent with previous reports in the literature, but confirmed here with a much larger sample. To understand the physical mechanisms governing this scatter, we search for correlations between the observed gradient and various stellar population properties at fixed mass. However, we find no evidence for a correlation with the galaxy properties we consider---including star-formation rates, sizes, star-formation rate surface densities, and star-formation rates per gravitational potential energy. We use the observed weakness of these correlations to provide material constraints for predicted intrinsic correlations from theoretical models.
\end{abstract}


\section{Introduction} \label{sec:intro}

As stars form and evolve in galaxies, they pollute their surroundings with metals. One might expect that the radial distribution of gas-phase metals (or, the abundance ratio of metals to non-metals, i.e., the metallicity) will follow the radial distribution of past generations of star-formation. At $1\,\lesssim\,z\,\lesssim\,3$, galaxies on average have negative radial gradients in star-formation rate surface density (e.g., \citealt{2016ApJ...828...27N}). In a simple closed-box with no radial transport of metals and no external sources acting to enrich or dilute the interstellar gas, local enrichment of the interstellar medium through star-formation should lead these galaxies towards negative radial gradients in gas-phase metallicity too---i.e., higher metallicities in their centers than in their outskirts. 

This description illustrates an important point, but is incomplete. Galaxies form and evolve in a complex ecosystem, with gas cycling in, out, and around galaxies on rapid timescales \citep{2015ARA&A..53...51S}. This cycle includes (1) stellar feedback driving metal-enriched winds out into the circumgalactic medium with the potential for re-accretion at later times \citep*{2017ARA&A..55..389T}, (2) metal-poor gas accretion onto galaxies from intergalactic filaments \citep{2005MNRAS.363....2K}, and (3) the rapid shuffling of metals through galaxy mergers \citep{2010ApJ...710L.156R}. These, and other actors, can effectively re-distribute metals on galaxy- and halo-scales. Indeed, by the present day, galaxies are estimated to have retained only $\sim$20$\%$ of the metals they produced over their lifetime \citep{2014ApJ...786...54P}---the rest presumably lost to the circum- and inter-galactic medium.

These processes are more prevalent at earlier times. At the peak of cosmic star-formation at $z\sim2$ \citep{2014ARA&A..52..415M}, 
%it is expected that 
the rates of star-formation of galaxies were 10 times higher \citep{2012ApJ...754L..29W}, the rates of accretion onto galaxies were $10-30$ higher \citep{2017ApJ...837..150S}, and the rates of galaxy-galaxy mergers were 10 times higher \citep{2017MNRAS.467.3083R} than they are today.

Numerical simulations reveal that these processes can {\emph{flatten}} gas-phase galaxy metallicity gradients---through e.g., star-formation and strong stellar feedback \citep{2013A&A...554A..47G, 2017MNRAS.466.4780M}, fountain flows \citep{2019MNRAS.490.4786G}, and mergers \citep{2010ApJ...710L.156R, 2011MNRAS.417..580P, 2012ApJ...746..108T}. As discussed at the outset, we expect that the simple continuous shedding of metals into the local interstellar medium through stellar evolution will continuously lead galaxies towards negative metallicity gradients. Observations of galaxies with flat or positive gradients (i.e., excursions from this simple expectation) lend insight into the prevalence and timescales of the processes that re-distribute metal-rich and metal-poor gas listed above. The demographics of gas-phase metallicity gradients over time (e.g., the population mean and scatter) provide an important benchmark for theoretical models of galaxy formation.

With the introduction of sensitive near-infrared spectrographs on the {\emph{Hubble Space Telescope}} and 10-m class ground-based facilities, the past decade has seen a rapid increase in observations of gas-phase metallicity gradients in high redshift galaxies \citep{2019A&ARv..27....3M}. Such measurements are now available for hundreds of galaxies at intermediate redshifts ($0<z<1$; e.g., \citealt{2018MNRAS.478.4293C, 2012ApJ...754...17F, 2014MNRAS.443.2695S, 2016ApJ...831..104G, 2019MNRAS.489..224P}) and high redshifts ($1 < z < 4$; e.g., \citealt{2010Natur.467..811C,2011ApJ...732L..14Y, 2012A&A...539A..93Q, 2012MNRAS.426..935S, Jones2013, 2015AJ....149..107J, 2016ApJ...827...74W, 2016ApJ...820...84L, 2017MNRAS.466..892M,  2017ApJ...837...89W, 2019ApJ...882...94W, 2020ApJ...900..183W, 2018ApJS..238...21F,  Curti_2019, 2020arXiv201015847G})---with samples that are large enough to draw meaningful conclusions about galaxy populations.

Up to $z\sim2$, high mass galaxies (M$_{*}\sim$10$^{10}$ - 10$^{11}$ M$_{\odot}$) tend to have slightly negative or flat metallicity gradients (e.g., \citealt{2016ApJ...827...74W, 2018ApJS..238...21F, 2018MNRAS.478.4293C}). Until recently, however, little information on the low mass galaxy population (M$_{*}<$10$^{10}$ M$_{\odot}$) above $z>1$ was available. The main limitation at these masses and redshifts is the small angular sizes---poor resolution leads to artificially flat metallicity gradients \citep{2013ApJ...767..106Y, 2017MNRAS.468.2140C, 2020MNRAS.495.3819A}. This poses a significant challenge for ground-based seeing-limited spectrographs, where the typical resolution is $\sim0\farcs5-1\farcs0$.

In the past few years, however, the low-mass population at these redshifts has started to be explored with large numbers (N $\sim$ 100) through deep galaxy surveys taking advantage of the magnification afforded through gravitational lensing. These include recent efforts with the {\emph{Hubble Space Telescope}}/Wide Field Camera 3 slitless spectrograph \citep{2017ApJ...837...89W, 2019ApJ...882...94W, 2020ApJ...900..183W} and the Very Large Telescope/KMOS integral field spectrograph \citep{Curti_2019}. These surveys reveal a zoo of flat, positive, and negative gradients---with a majority of galaxies having flat gradients. These results are in marked contrast with today's star-forming galaxies, of which the majority show declining gas-phase metallicity gradients (e.g., \citealt{2017MNRAS.469..151B}).

In this paper, we use deep {\emph{Hubble}} slitless grism spectroscopy from the Wide Field Camera 3 to study the metallicity gradients of \samplesize{} galaxies at \redshiftrange{} over a stellar mass range of \massrange{}---effectively doubling the low mass sample size at these redshifts. Our goal is to provide a statistical understanding of the gas phase metallicity gradients at high redshift, and to examine the population scatter as a function of stellar mass.

Our outline for this paper is as follows. In \S\ref{sec:observations_fits_sample}, we detail the observations, data reduction, and sample used in this paper. In \S\ref{sec:measurements}, we describe the measurements of the physical properties of our sample, including the metallicity gradients. Next, in \S\ref{sec:metal_gradients}, we examine the gas phase metallicity gradients as a function of galaxy stellar mass and assess the intrinsic scatter of the population. We also explore correlations between the metallicity gradient and stellar population properties. We use these results to place constraints on predictions of correlation strengths from theoretical models. In \S\ref{sec:discussion}, we discuss our findings and, finally, in \S\ref{sec:conclusions} we summarize the results of the paper.

Throughout this paper, we adopt a $\Lambda$CDM cosmology with the Planck 2015 cosmological parameters (h, $\Omega_m$, $\Omega_\lambda$) = (0.67, 0.31, 0.69). For relevant derived quantities (e.g., stellar masses, star-formation rates), we assume a \citet{2003PASP..115..763C} initial mass function.

\section{Observations, Data Reduction, and Sample Selection}\label{sec:observations_fits_sample}

In this section, we describe the observations (\S\ref{sec:observations}), data reduction/spectral fitting (\S\ref{sec:fits}), and sample selection (\S\ref{sec:sample}) used in this paper.

\subsection{Hubble Slitless Spectroscopy}\label{sec:observations}
The {\emph{Hubble}} Wide Field Camera 3 ({\emph{HST}}/WFC3) slitless spectra used in this paper are taken from a collection of programs. The root program is the Cycle 23 CANDELS Lyman-$\alpha$ Emission at Reionization survey (CLEAR; GO-14227, PI Papovich). The CLEAR footprint covers 12 pointings in the GOODS-S deep and GOODS-N deep CANDELS fields \citep{2011ApJS..197...35G, 2011ApJS..197...36K}. Each pointing includes 10 to 12-orbit depth (10 for GOODS-N and 12 for GOODS-S) {\emph{HST}}/WFC3 G102 grism spectroscopy and companion {\emph{HST}}/WFC3 F105W direct imaging. Each pointing is observed at 3 orients, separated by $>$10 degrees---helping relieve source confusion in the grism spectra. A primary motivation for the CLEAR survey was to constrain the evolution of the distribution function of Lyman-$\alpha$ emission at $z>6.5$. Such constraints from CLEAR data will be presented in a forthcoming paper (Jung et al. in prep). Previous work has used the CLEAR spectra to study the metallicities, ages, and formation histories of massive high redshift galaxies (\citealt{2019ApJ...870..133E, 2020ApJ...898..171E}), and to appraise Paschen-$\beta$ as a star-formation rate indicator in low redshift galaxies \citep{2020arXiv200900617C}.

Given their location in the well-studied GOODS-S and GOODS-N fields, the CLEAR observations are supported with extensive ancillary photometry spanning the ultraviolet to the near-infrared. We use an augmented version of the v4.1 3D-HST GOODS-S and GOODS-N photometric catalogs \citep{2014ApJS..214...24S}---adding $Y$-band photometry measured from the CLEAR F105W imaging and archival {\emph{HST}}/WFC3 F098M and/or F105W imaging. $Y$-band imaging is available over the majority of the GOODS-S and GOODS-N fields, and the entire CLEAR footprint. The new $Y$-band photometry is measured in a manner that is consistent with the existing 3D-HST photometric catalog (see details in \citealt{2019ApJ...870..133E} and \citealt{2014ApJS..214...24S}). We use the python version of {\tt{Eazy}}\footnote{https://github.com/gbrammer/eazy-py} \citep{2008ApJ...686.1503B} to re-derive the photometric zeropoints of the augmented catalog.

To extend the spectral coverage and maximize the depth of our grism observations, we query the Mikulski Archive for Space Telescopes (MAST) for all publicly-available {\emph{HST}}/WFC3 G102 and G141 grism observations (and their associated direct imaging) that overlap with the root CLEAR footprint. In doing so, we retrieve a total of 52 additional orbits with the WFC3/G102 grism and 76 orbits with the redder WFC3/G141 grism. Hereafter, we refer to the full collection of grism spectra as `CLEAR+'.

The archival G141 observations were acquired by the following programs: GO-11600 (`AGHAST'; PI Weiner), GO-12461 (`SN COLFAX', PI Reiss), GO-13871 (PI Oesch), GO/DD-11359 (`ERS', PI: O'Connell), GO-12099 (`GEORGE, PRIMO', PI Reiss), and GO-12177 (`3D-HST', PI van Dokkum). The archival G102 observations are from: GO-13420 (PI Barro), GO/DD-11359 (`ERS', PI O'Connell), and GO-13779 (`FIGS', PI Malhotra).

\begin{figure}
    \centering
    \includegraphics[width = \columnwidth]{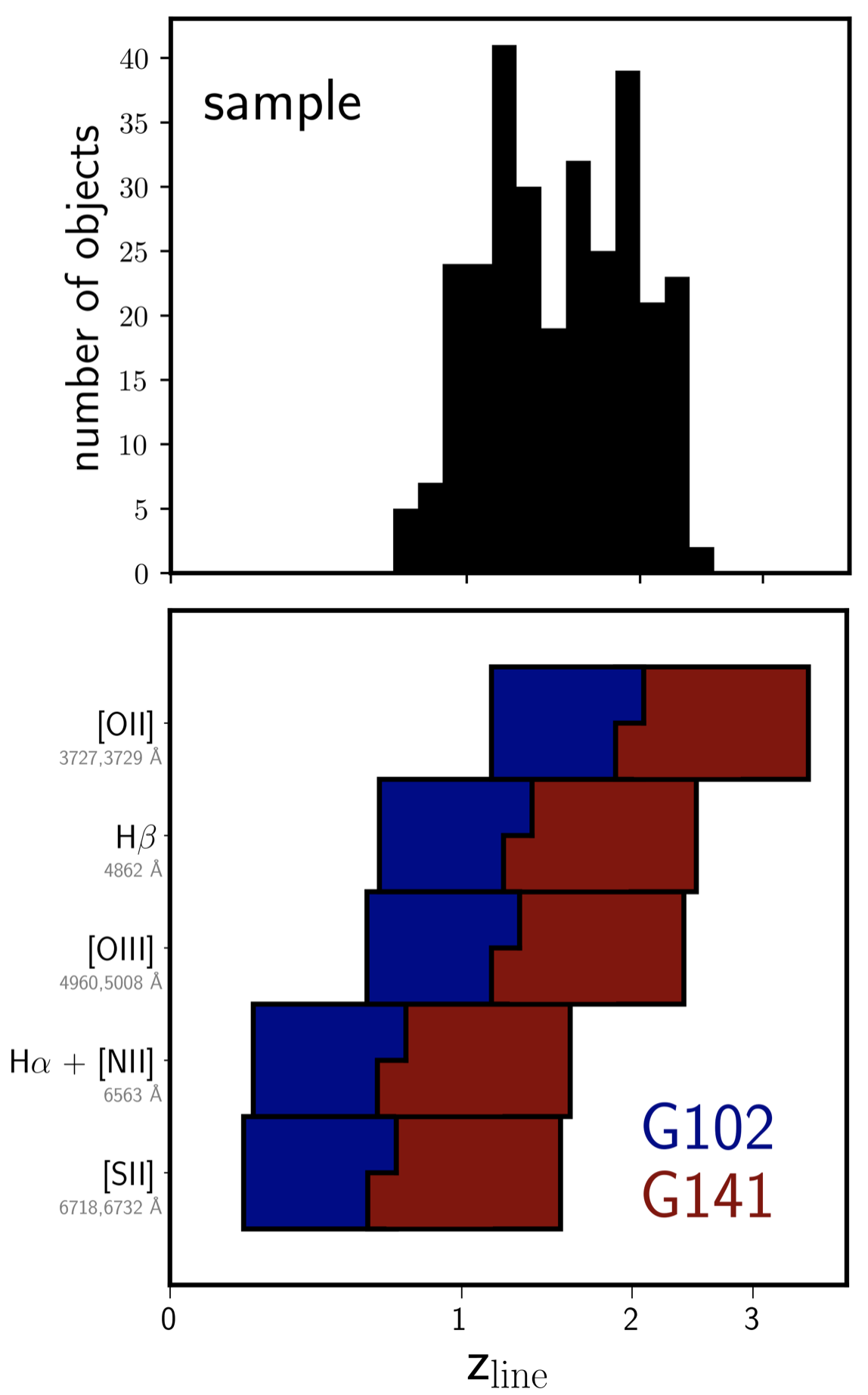}
    \caption{The redshift windows in which optical strong lines are accessible with the {\emph{HST}}/WFC3 G102 (blue) and G141 (red) grism spectrographs are shown (bottom panel). The galaxies used in this paper are selected on 5$\sigma$ integrated detections in at least two lines in the R$_{\text{23}}$ complex: [O\ii], [O\iii], and H$\beta$. This criterion establishes lower and upper redshift windows of $z$ = 0.6 and 2.6, respectively.  The redshift distribution of our sample is shown in the top panel. }
    \label{fig:lines_grism}
\end{figure}

In Figure \ref{fig:lines_grism}, we show the redshift windows over which several strong rest-frame optical lines are accessible with the G102 (0.80--1.15$\,\mu$m, $R\sim210$) and G141 (1.08--1.70$\,\mu$m,  $R\sim130$) grisms---illustrating a notable gain with joint G102 + G141 coverage over either of them individually. By themselves, the WFC3 grisms offer only narrow redshift windows in which these key metallicity diagnostics ([S\ii], H$\alpha$ + [N\ii], [O\iii], H$\beta$, [O\ii]; \citealt{2019A&ARv..27....3M}) are {\emph{simultaneously}} visible. The full R$_{\text{23}}$ complex (H$\beta$, [O\iii], and [O\ii]) can be accessed over just $1.2\,<\,z\,<1.3$ with the G102 grism and $2.0\,<\,z\,<2.4$ with the G141 grism. With joint WFC3/G102+G141 spectral coverage, the same redshift windows are significantly wider---the full R$_{\text{23}}$ complex can be accessed over $1.2\,<\,z\,<2.4$ when both grisms are employed. Limiting to any 2 lines of the R$_{\text{23}}$ complex, the redshift window increases to \redshiftrange{} with both grisms employed. Furthermore, the spectral overlap of the G102 and G141 grisms leads to continuous coverage over the full spectral range.

\subsection{Grism Data Reduction and Spectral Extractions}\label{sec:fits}

To process the set of G102 + G141 grism observations described in the previous subsection, we use the grism redshift and line analysis software {\tt{Grizli}}\footnote{https://github.com/gbrammer/grizli}\citep{2019ascl.soft05001B}. {\tt Grizli} performs full end-to-end processing of {\emph{HST}} imaging and slitless spectroscopic datasets, including: retrieving and pre-processing the raw observations, astrometric alignment, modeling contamination from overlapping spectra, extracting 1D and 2D spectra, fitting full continuum + emission line models, and generating emission line maps. 

First, we use {\tt{Grizli}} to retrieve and pre-process the observations described in \S \ref{sec:observations} from the MAST archive. The raw WFC3 data are reprocessed with the \texttt{calwf3} pipeline with corrections for variable sky backgrounds as described by \citet{2016wfc..rept...16B}. Cosmic rays and hot pixels not flagged by the pipeline are identified with the AstroDrizzle software \citep{2012drzp.book.....G}.  As in \citet{2016ApJS..225...27M}, the grism exposures are flat-fielded using the F105W and F140W calibration images for the G102 and G141 grisms, respectively, and grism sky subtraction is performed using the ``Master Sky'' provided in \citet{2015wfc..rept...17B}.  Relative astrometric corrections are applied to the processed data through an alignment to the deeper F140W {\emph{HST}} mosaic galaxy catalog from the 3D-HST survey \citep{2014ApJS..214...24S}.

A contamination model of each CLEAR+ pointing is created using a forward-model of the {\emph{HST}} Y-band full-field mosaic. For each spectrum, this model is used to subtract the contamination from adjacent spectra. This is an iterative process. The first pass contamination model is created for all objects in the $Y$-band segmentation map brighter than m$_{\text{F105W}}$ $=$ 25, assuming spectra that are flat in units of $f_\lambda$ flux density and normalized at F105W given the image segment. Next, a refined continuum model is created for objects brighter than m$_{\text{F105W}}$ $=$ 24 by fitting 3rd-order polynomials to the spectrum of each source after subtracting the models of contaminating sources. These steps result in a full contamination model of the detector for each visit of each observational program.

We use {\tt{Grizli}} to extract the 2D grism spectra (i.e., ``beams'') of objects in the field of view to a limiting magnitude of m$_{\text{F105W}}$ $<$ 25. This is performed for each grism visit of each object. The extractions carry the full description of the WFC3 detector and the contamination model. The grism exposure times of the CLEAR+ objects range from $2-30$ hours in G102 and $2-10$ hours in G141. In total, 6048 objects are extracted from the CLEAR+ data. Of these, 533 have sole coverage with the G102 grism, 808 have sole coverage with the G141 grism, and 4707 have joint coverage with both grisms.

Redshift fits are carried out using the grism spectra and available multiwavelength photometry. The spectra are scaled to the photometry using a 1st-order polynomial correction. A basis set of template  Flexible Stellar Population Synthesis models ({\tt{FSPS}}; \citealt{2009ApJ...699..486C, 2010ApJ...712..833C}), including emission line complexes with fixed ratios, are used in the fit. The {\tt{FSPS}} templates are constructed to span a broad diversity of galaxy types following the methodology described by \citet{2007AJ....133..734B} and \citet{2008ApJ...686.1503B}. At a given trial redshift, the redshifted templates are both convolved with the photometric filter bandpasses and projected to the space of each extracted 2D spectral ``beam'' using the direct $Y$-band image to define the spatial morphology. This approach takes into account the unique morphological broadening of each galaxy due to the limited spectral resolution of the grism.  A final model is determined from a non-negative linear combination of the model template spectra and goodness of fit is determined from the combined $\chi^2$ of all photometry and 2D spectral pixels using the uncertainties from the photometric catalogs and exposure-level noise model, respectively.  The ``best'' redshift is taken to be where this $\chi^2$ is minimized across a grid of trial redshifts from $0 < z < 12$.

Emission line fluxes are fit at the best-fit redshift using the {\tt{FSPS}} basis templates described above, but now including separate components for each line species without fixing their line ratios\footnote{The [\ion{O}{3}]$\lambda\lambda$4959+5007 and [\ion{S}{2}]$\lambda\lambda$6717+6731 doublets are fit as single components with fixed line ratios 1:2.98 and 1:1, respectively. The H$\alpha$+[\ion{N}{2}] complex is fit as a single component at the wavelength of H$\alpha$.}. These fits are carried out using the forward-modelling technique described above.

The {\tt{Grizli}}-derived redshifts and emission line fluxes are consistent with previous measurements from the 3D-HST survey \citep{2016ApJS..225...27M}---their differences divided by the sum of their uncertainties have a near standard normal distribution in both cases.  Comparing to ground-based spectroscopic measurements of sources in the CLEAR fields, we find an overall redshift precision of $\sigma_\mathrm{NMAD}\,=\,0.0024$ in $\Delta z/(1+z)$.

Emission line maps are created for several lines (notably the strong lines listed in Figure \ref{fig:lines_grism}) by drizzling the contamination- and continuum-subtracted 2D spectral beams using the astrometry of the slitless exposures projected along the spectral trace to the wavelength of the redshifted line center. The line maps are created with a  pixel scale of 0\farcs1. The uncertainties on the line maps are calculated using the drizzle weights from the constituent beam pixels. Example line maps are shown in the top panels of Figure \ref{fig:example_metallicity_maps}.  
%
% Casey added a comment (response to Gabe) about adding a discussion of the errors on the line maps. 

\begin{figure*}
    \centering
    \includegraphics[width = \textwidth]{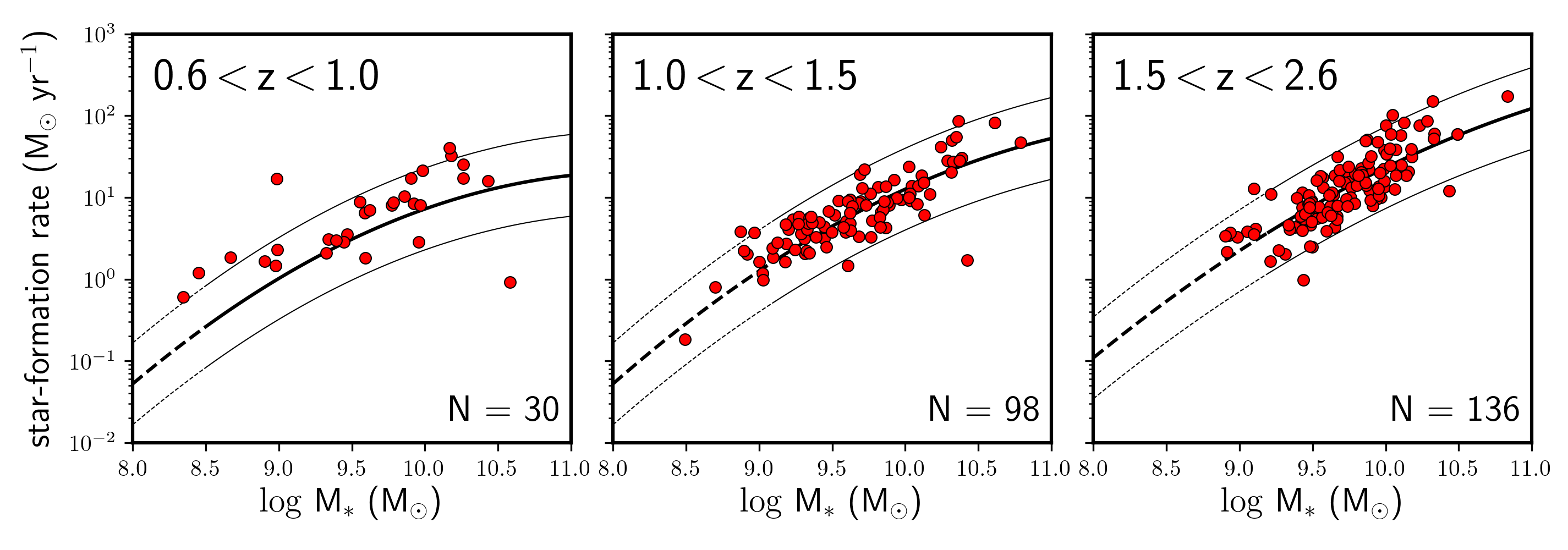}
    \caption{The star-formation rate versus stellar mass diagram for the galaxies in our sample is shown. The sample is broken into three redshift ranges, each spanning $\sim$2 Gyr in cosmic time. Thick black lines show the star-formation mass sequence (SFMS) at each respective epoch, taken from \citet{2014ApJ...795..104W}. The thin black lines show 1/2-dex above and below the SFMS. The number of galaxies in each panel is listed in the lower right. The galaxies in this sample lie along the SFMS.}
    \label{fig:m_sfr}
\end{figure*}

\subsection{Sample Selection}\label{sec:sample}

To choose galaxies we can reasonably derive gas-phase metallicity maps for, we select objects with a 5$\sigma$ integrated detection in at least 2 of the strong lines in the R$_{\text{23}}$ complex: [O\iii]$\lambda5007,4958$, $[\text{O\ii}]\lambda3727$, and $\text{H}\beta$.  This ensures access to at least one set of the metallicity-sensitive line ratios
\begin{align}
\text{O}_{32} &\equiv \log \left([\text{O\iii}]\lambda5007 / [\text{O\ii}]\lambda3727\right) \nonumber \\
\text{R}_{23} &\equiv \log \left(\left([\text{O\iii}]\lambda5007,4958 + [\text{O\ii}]\lambda3727\right)/\text{H}\beta\right) \nonumber \\
\text{R}_3  &\equiv \log \left([\text{O\iii}]\lambda5007,4958 /\text{H}\beta\right) \nonumber \\
\text{R}_2  &\equiv \log \left([\text{O\ii}]\lambda3727/\text{H}\beta\right) \nonumber
\end{align}

\noindent This selection leads to a total of 486 galaxies. Of these, 19$\%$ (N = 86) have spectral coverage of the full R$_{\text{23}}$ complex, 39$\%$ (N = 190) have coverage of only the R$_{\text{3}}$ complex, less than 2$\%$ have coverage of only the R$_{\text{2}}$ complex (N = 9), and 41$\%$ have coverage of only the O$_{32}$ complex (N = 201).

We match the CLEAR sample to the {\emph{Chandra}} Deep Field X-ray point source catalogs \citep{2016ApJS..224...15X, 2017ApJS..228....2L} and remove galaxies harboring X-ray bright Active Galactic Nuclei (AGN). AGN ionization is not accounted for in the photoionization models. It will act to increase the O$_{32}$ ratio and artificially decrease the inferred metallicities in the centers of galaxies. This selection removes 28 galaxies from the sample.

As discussed in the following section, the sample is further culled through a selection on the radial extent of the detected metallicity signal. This leaves \samplesize{} galaxies in our final science sample. The redshift distribution of this sample is shown in the top panel of Figure \ref{fig:lines_grism}. The selection criterion constrains the redshifts to \redshiftrange{}. The majority of the sample are at the high redshift end of this range---$11\%$, $37\%$, and 52$\%$ of the sample are at $0.6\,<\,z\,<\,1.0$, $1.0\,<\,z\,<\,1.5$, and $1.5\,<\,z\,<\,2.6$, respectively.

\section{Measurements of Physical Properties}\label{sec:measurements}

In this section, we describe the measurements of the physical properties of the CLEAR+ sample. In \S\ref{sec:sed_properties}, we discuss fits to their stellar masses, star-formation rates, dust extinctions, and sizes. In \S\ref{sec:metallicity_maps}, we derive {\emph{HST}}-resolution metallicity maps of the CLEAR+ galaxies using pixel-by-pixel fits to the emission line maps, and fit the radial gradients of the gas-phase metallicity.

\subsection{Stellar Mass, Star-Formation Rates, Dust Extinction, and Sizes}\label{sec:sed_properties}

Stellar masses, star-formation rates, and dust extinctions are measured from the ancillary multi-wavelength photometry of the full CLEAR+ sample using {\tt{Eazy-py}}. The fits are carried out with the `fsps\_QSF\_12\_v3' SED template set available in the {\tt{Eazy}} library. These templates are created using {\tt{FSPS}}, described in \S\ref{sec:fits}, and assume a \citet{2003PASP..115..763C} initial mass function (IMF).

The circularized effective radii (R$_{\text{eff}}$) of the CLEAR+ galaxies were measured in \citet{2012ApJS..203...24V}. We adopt R$_{\text{eff}}$ as measured from the CANDELS {\emph{HST}}/WFC3 F125W+F160W imaging, with the redshift-dependent correction outlined in \citet{2012ApJS..203...24V}.

UV + IR-derived star-formation rates are taken from \citet{2014ApJ...795..104W}. These are measured using a conversion from total UV + IR luminosity to total star-formation rate \citep{2005ApJ...625...23B}, assuming a \citet{2003PASP..115..763C} IMF. A template-based \citep{2002ApJ...576..159D} conversion is used to translate the observed \emph{Spitzer}/MIPS 24 $\mu$m flux density of each galaxy to its total IR luminosity. The total UV luminosity is linearly scaled from the rest-frame 2800 \AA\,luminosity derived from the best-fit SED template. UV + IR star-formation rates are available for $\sim$83\% of our sample. For galaxies where it is not available, we use the SED-derived star-formation rates described above.

\begin{figure*}
    \centering
    \includegraphics[width =\textwidth]{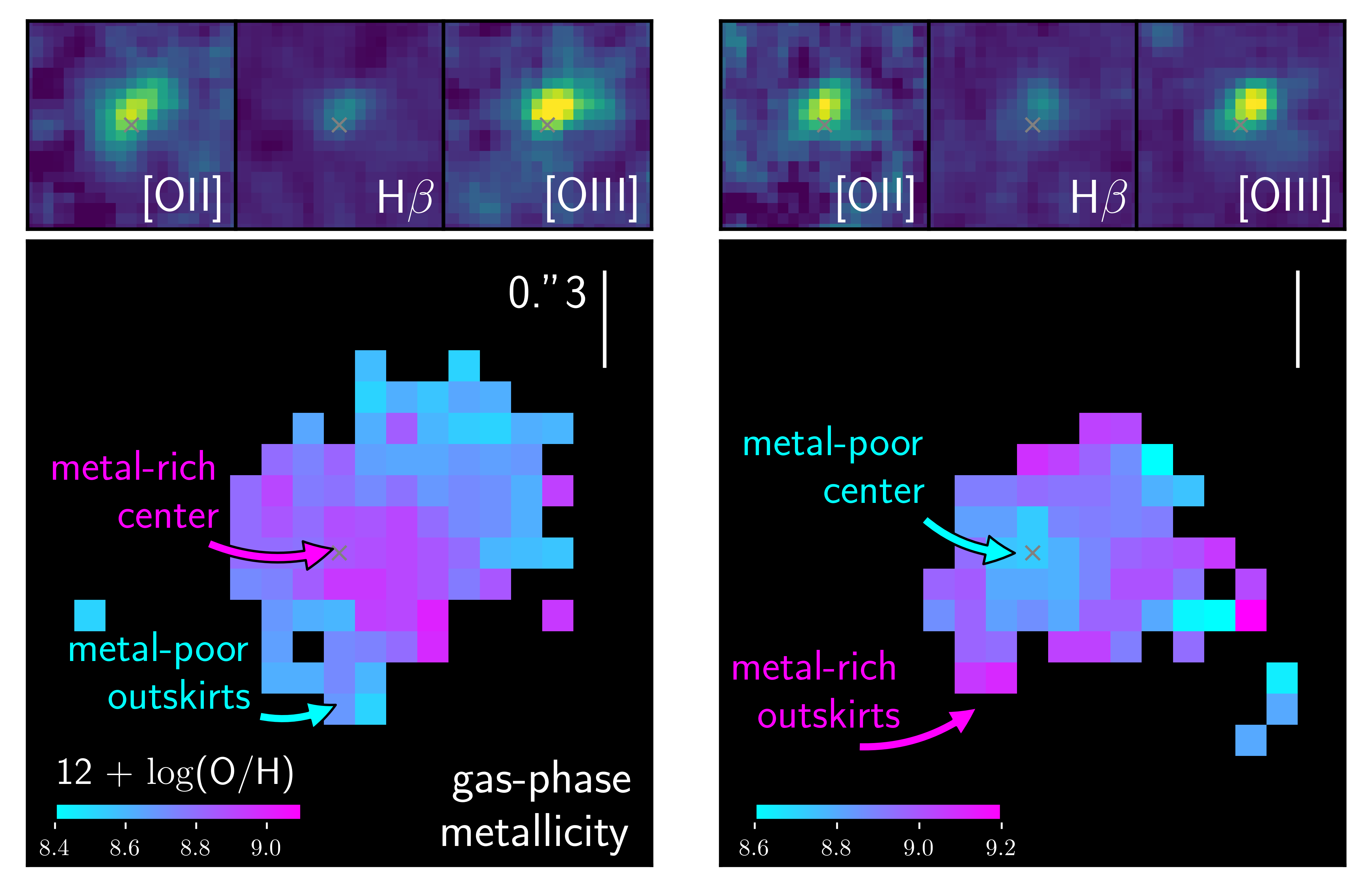}
    \caption{Example emission line and gas-phase metallicity maps are shown for 2 galaxies in the CLEAR sample. Emission line maps of [O\ii], H$\beta$, and [O\iii] are shown in the top row. The best-fit metallicity map is shown in the bottom row. The examples include a galaxy that is more metal-rich in its center than in its outskirts (i.e., a galaxy with a negative metallicity gradient; left) and a galaxy that is more metal-poor in its center than in its outskirts (i.e., a galaxy with an positive metallicity gradient; right). A grey x marks the center of each galaxy and a 0.$\arcsec$3 scale is included for reference.}
    \label{fig:example_metallicity_maps}
\end{figure*}

\begin{figure*}
    \centering
    \includegraphics[width =\textwidth]{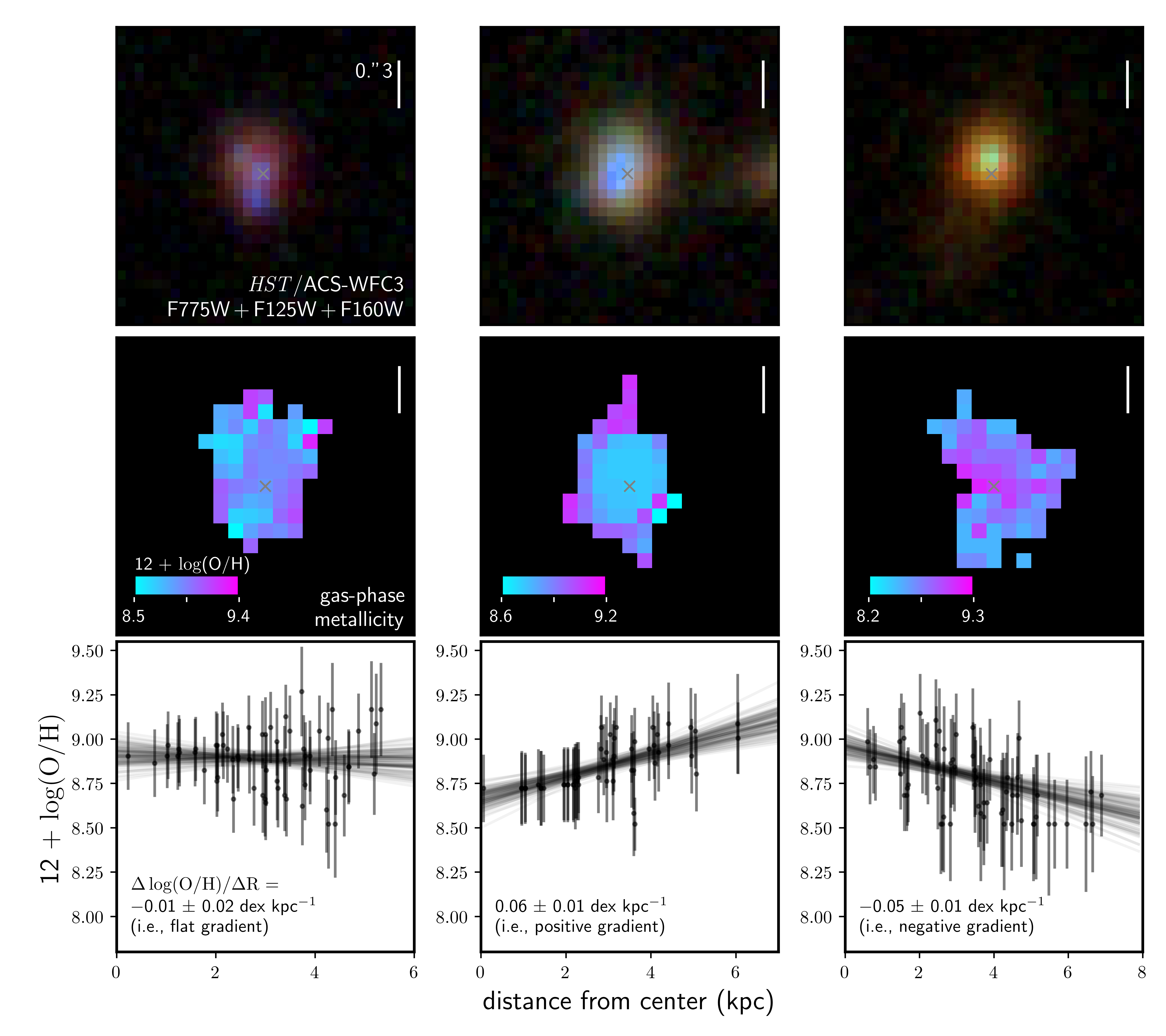}
    \caption{Example fits to the radial gas-phase metallicity profile for 3 galaxies in the CLEAR sample are shown, including: a galaxy with a flat metallicity gradient (left column), a galaxy with a positive gradient (middle column), and a galaxy with a negative gradient (right column). An HST/ACS-WFC3 color map is shown in the top row. The gas-phase metallicity map is shown in the middle row and the radial metallicity profile is shown in the bottom row. In the top and middle rows, a grey square is shown to mark the center of the galaxy, as defined by the center-of-light of the {\emph{HST}}/WFC3 F105W image, and a white bar extending 0.$\arcsec$3 is shown for scale. Random draws from the posteriors of the linear fits to the metallicity profile are shown in black in the bottom row.}
    \label{fig:example_metallicity_gradients}
\end{figure*}

In Figure \ref{fig:m_sfr}, we plot the star-formation as a function of stellar mass for our sample of \samplesize{} galaxies. We also include the star-formation mass sequence (SFMS) fits from \citet{2014ApJ...795..104W}. The galaxies in our sample lie along the SFMS at their respective epochs.

\begin{deluxetable*}{c|c|c|c|c|c|c|c}
    \setlength{\tabcolsep}{15pt} 
    \centering
    \tablecaption{Metallicity Gradients of CLEAR+ Sample (0.6 $<$ z $<$ 2.6)}
    \tablehead{Field & ID & R.A. & DEC. & z & $\log$ M$_*$  & slope, $m$       & intercept, $b$ \\
                     &    & J2000   &  J2000   &   & (M$_{\odot}$) & (dex kpc$^{-1}$) & (dex) }
        \startdata
        GOODS-N & 11883 & 189.152027 & +62.2008193 & 0.97 & 8.99 &\hspace{0.23cm}0.055 $\pm$ 0.025 & 8.853 $\pm$ 0.052\\
        GOODS-N & 11502 & 189.260943 & +62.1991988 & 1.99 & 9.49 &\hspace{0.23cm}0.093 $\pm$ 0.015 & 8.763 $\pm$ 0.048\\
        GOODS-N & 14506 & 189.1121534 & +62.2133287 & 1.67 & 9.76 &\hspace{0.23cm}0.010 $\pm$ 0.013 & 8.807 $\pm$ 0.050\\
        GOODS-N & 17927 & 189.103116 & +62.2301769 & 1.35 & 9.34 &$-$0.013 $\pm$ 0.045 & 8.690 $\pm$ 0.098\\
        GOODS-N & 10964 & 189.1430768 & +62.1966815 & 1.51 & 9.31 &\hspace{0.23cm}0.136 $\pm$ 0.032 & 8.437 $\pm$ 0.071\\
        GOODS-N & 16500 & 189.1317551 & +62.2235485 & 1.79 & 9.27 &\hspace{0.23cm}0.140 $\pm$ 0.049 & 8.399 $\pm$ 0.104\\
        GOODS-N & 15474 & 189.1757067 & +62.2180897 & 2.00 & 9.84 &\hspace{0.23cm}0.087 $\pm$ 0.024 & 8.744 $\pm$ 0.068\\
        GOODS-N & 30204 & 189.3194715 & +62.2925881 & 1.15 & 10.70 &\hspace{0.23cm}0.114 $\pm$ 0.025 & 8.651 $\pm$ 0.076\\
        GOODS-S & 26406 & 53.1450689 & $-$27.7894258 & 1.32 & 9.86 &$-$0.021 $\pm$ 0.008 & 8.975 $\pm$ 0.028\\
        GOODS-S & 29460 & 53.1833594 & $-$27.7761358 & 1.55 & 9.19 &$-$0.030 $\pm$ 0.017 & 8.712 $\pm$ 0.052\\
        GOODS-S & 45789 & 53.2561716 & $-$27.6950420 & 0.98 & 10.61 &\hspace{0.23cm}0.102 $\pm$ 0.015 & 8.486 $\pm$ 0.042\\
        GOODS-S & 26387 & 53.1738637 & $-$27.7884904 & 1.67 & 9.67 &\hspace{0.23cm}0.083 $\pm$ 0.019 & 8.782 $\pm$ 0.056\\
        GOODS-S & 26698 & 53.1640939 & $-$27.7872909 & 1.10 & 9.09 &\hspace{0.23cm}0.176 $\pm$ 0.025 & 8.345 $\pm$ 0.069\\
        GOODS-S & 40759 & 53.0569153 & $-$27.7203035 & 1.47 & 9.92 &\hspace{0.23cm}0.006 $\pm$ 0.010 & 9.052 $\pm$ 0.040\\
        GOODS-S & 36182 & 53.1602535 & $-$27.7432660 & 0.96 & 8.67 &\hspace{0.23cm}0.073 $\pm$ 0.040 & 8.653 $\pm$ 0.094\\
        GOODS-S & 40108 & 53.1680806 & $-$27.7235106 & 1.25 & 9.69 &$-$0.001 $\pm$ 0.009 & 8.910 $\pm$ 0.033\\
        ... & ... & ...& ... & ...  & ...  &...  & ... \\
        \enddata
\tablecomments{Linear fits to the radial profiles of the gas-phase metallicity for a subset of the galaxies used in this paper. The fits take the simple form $\log$(O/H)($R$) = $m\,\times\,R\,+\,b$, where $R$ is the projected radial distance from the galaxy center. The IDs are matched to the 3D-HST photometric catalog \citep{2014ApJS..214...24S}. The full data table will be made available online alongside publication.}
\label{tab:partial_table}
\end{deluxetable*}

\begin{deluxetable}{c|c|c|c}
    \setlength{\tabcolsep}{15pt} 
    \centering
    \tablecaption{Metallicity Gradients of Population Stacks (0.6 $<$ z $<$ 2.6)}
    \tablehead{
         $\log$ M$_*$& N & slope, $m$ & intercept, $b$ \\
         (M$_{\odot}$) & &(dex kpc$^{-1}$) & (dex)}
    \startdata
8.5 - 9.0          & 15&0.059 $\pm$ 0.024 & 8.651 $\pm$ 0.040 \\
9.0 - 9.5          & 71&0.052 $\pm$ 0.003 & 8.721 $\pm$ 0.006 \\
$\,\,\,$9.5 - 10.0 & 119&0.019 $\pm$ 0.002 & 8.823 $\pm$ 0.006 \\
10.0 - 10.5        & 63&0.005 $\pm$ 0.002 & 8.928 $\pm$ 0.006 \\
10.5 - 11.0        & 24&0.022 $\pm$ 0.013 & 8.909 $\pm$ 0.043 \\
    \enddata
    \tablecomments{Linear fits to the radial profiles of gas-phase metallicity for stacks of CLEAR galaxies binned by stellar mass. These are shown in the right panel of Figure \ref{fig:gradients_our_sample}. The number of galaxies contributing to each stack (N) is listed.}
    \label{tab:stack_table}
\end{deluxetable}

\subsection{Metallicity Maps}\label{sec:metallicity_maps}

We carry out pixel-by-pixel fits of metallicity and ionization parameter to the grism-derived emission line maps using the bayesian photoionization fitting code {\tt{IZI}}\footnote{https://users.obs.carnegiescience.edu/gblancm/izi/} \citep{2015ApJ...798...99B}. {\tt{IZI}} uses a grid of outputs from a user-specified photo-ionization model to fit observed dust-corrected emission line fluxes. We dust-correct the observed emission line maps using the {\tt{Eazy}}-derived $A_V$ value of each galaxy, assuming a \citet{2000ApJ...533..682C} extinction law. We do not account for differences between stellar and nebular extinction at these redshifts \citep{2014ApJ...788...86P} and assume no radial dependence on $A_V$. We test the biases introduced by the latter assumption using a suite of simulated profiles. These simulated profiles span a range of intrinsic gas-phase metallicity gradients, with a fixed central $A_V\,=\,1.5$ (typical of the galaxies in our sample), and an intrinsic gradient in $A_V$ of $-0.08$ dex kpc$^{-1}$. The latter is chosen to match the broadband-derived measurements of high mass galaxies ($\,>\,10^{10}$ M$_{\odot}$) at $z\,=\,2$ \citep{2018ApJ...859...56T}. Using the same measurement techniques we use for the real galaxies in our sample (outlined below), we find that discluding a dust gradient does not significantly bias the recovered metallicity gradients---there is a $\sim5-30\%$ bias depending on the intrinsic metallicity gradient and the metallicity diagnostics available. Furthermore, the dust gradients in low mass galaxies ($\,<\,10^{10}$ M$_{\odot}$) at the redshifts of our sample have been measured to be relatively flat on average \citep{2016ApJ...817L...9N}. Low mass galaxies comprise the majority of our sample. Taken as a whole, and without uniform empirical constraints for our whole sample, we consider the choice of a flat dust gradient justified.

We use the {\tt{MAPPINGS-IV}} photo-ionization models \citep{2013ApJS..208...10D}, with the parameter describing the distribution of electron energies $\kappa$ set to 20. The  {\tt{MAPPINGS-IV}} models adopt a functional dependence of the nitrogen-to-oxygen ratio and the oxygen abundance, based on observations of local galaxies \citep{1998AJ....116.2805V}. Because the gradient is a relative measure, the results presented herein are less sensitive to the normalization-differences between photo-ionization libraries. A comparison between the predictions of the photo-ionization models and observed line ratios from the CLEAR+ sample appears elsewhere (Papovich et al.\ in prep).

For each object, we create a segmentation map where the [O\ii] line map (or H$\beta$, if [O\ii] is not observed) exceeds a S/N of 1. While this S/N threshold is relatively low, the S/N of the H$\beta$ and [O\ii] line fluxes are generally the lowest of the line fluxes that are used in the photonionization fit. Furthermore, the metallicity gradients (described later) are fit using a large number of pixels and are not generally susceptible to low-S/N outliers. Pixels inside this segmentation region are fit with {\tt{IZI}}, and pixels outside are masked. For each $0.\arcsec1\times0.\arcsec1$ pixel in the unmasked CLEAR line maps, a posterior inference of the metallicity and ionization parameter is recovered. {\tt{IZI}} is flexible---it allows for upper limits as well as line sums (from e.g., unresolved line complexes in the grism). We provide {\tt{IZI}} the strong emission lines and line sums listed in Figure \ref{fig:lines_grism}: [S\ii], H$\alpha$ + [N\ii], [O\iii]$\lambda$5007,4958, H$\beta$, and [O\ii]$\lambda$3727.

The R$_{\text{2}}$, R$_{\text{3}}$, and R$_{\text{23}}$ ratios have degenerate solutions with metallicity---a low metallicity branch and a high metallicity branch. An ancillary measurement of other line ratios e.g., O$_{32}$, can help break this degeneracy (see C. Papovich et al., in prep, for the application to the CLEAR+ data), but high S/N detections of the [O\ii] line are not always available. Without such a constraint, the branch favored by the posterior is generally a strong function of the metallicity prior adopted. To avoid internal ``branch-switching" in our fits---i.e., a subset of galaxy pixels marginally favoring the lower branch and the remaining pixels marginally favoring the upper branch---we set a prior that favors the high-metallicity branch of the R$_{\text{2}}$, R$_{\text{3}}$, and R$_{\text{23}}$ {\tt{MAPPINGS-IV}} models. Specifically, we adopt a top-hat prior over 12 + $\log$ O/H = [8.5, 9.5]. This choice is supported by the locus of the global gas-phase mass-metallicity relation (MZR) at the extreme ends of our sample---the highest redshifts and lowest masses, where for our sample very few objects have integrated gas-phase Oxygen abundances that fall on the lower branch derived from the integrated R$_\mathrm{23}$ and O$_\mathrm{32}$ emission lines (C.~Papovich et al.\ 2020, in prep).  This is consistent with other measurements of galaxies at $z\,\sim\,2.3$, which indicate that the integrated MZR relation intercepts 12 + $\log$ O/H $\sim$ 8.4 at M$_{*}\,\sim\,10^{9.5}$ M$_{\odot}$ and increases with increasing mass and decreasing redshift \citep{2018ApJ...858...99S, 2020arXiv200907292S}.

Example emission line maps, and derived metallicity maps, are shown in Figure \ref{fig:example_metallicity_maps}. We adopt the luminosity-weighted center of the direct F105W image as the center of each galaxy, and measure the radial position and gas-phase metallicity of each pixel. We do not de-project the galactic coordinates, as the emission line kinematics of galaxies at this redshift indicate that the emission line structure is rarely disky \citep{2012ApJ...758..106K, 2016ApJ...830...14S, 2017ApJ...843...46S}. As such, a de-projection using e.g., the continuum axis ratio, has uncertain meaning. Finally, we remove galaxies from the sample if their metallicity is not recovered beyond 0.$\arcsec$3 arcsec. This ensures that the radial profiles in our final sample extend at least $\sim$2 {\emph{HST}}/WFC3 PSF FWHM resolution elements.

To derive the metallicity gradients, we perform a least-squares fit to the radial metallicity profile of each galaxy using a simple line of the form $\log$(O/H)($R$) = $m\,\times\,R\,+\,b$, where $R$ is the projected radial distance from the galaxy center, $m$ is the metallicity gradient, and $b$ is the central metallicity. Example fits are shown in Figure \ref{fig:example_metallicity_gradients}. A truncated table of the best-fit parameters is shown in Table \ref{tab:partial_table}.

Finally, we split our sample into sub-populations binned by 0.5 dex intervals of stellar mass. In each bin, we perform a median stack of the radial profiles of the constituent galaxies. In each radial bin of each stack, we require at least 10 individual galaxies with metallicity information. We bootstrap re-sample the stacked profiles and measure the median and uncertainty on the median in each bin. We fit the stacked profiles using the linear fit described above and report the best-fit parameters in Table \ref{tab:stack_table}.

\section{Metallicity Gradients}\label{sec:metal_gradients}

\begin{figure*}
    \centering
    \includegraphics[width=\textwidth]{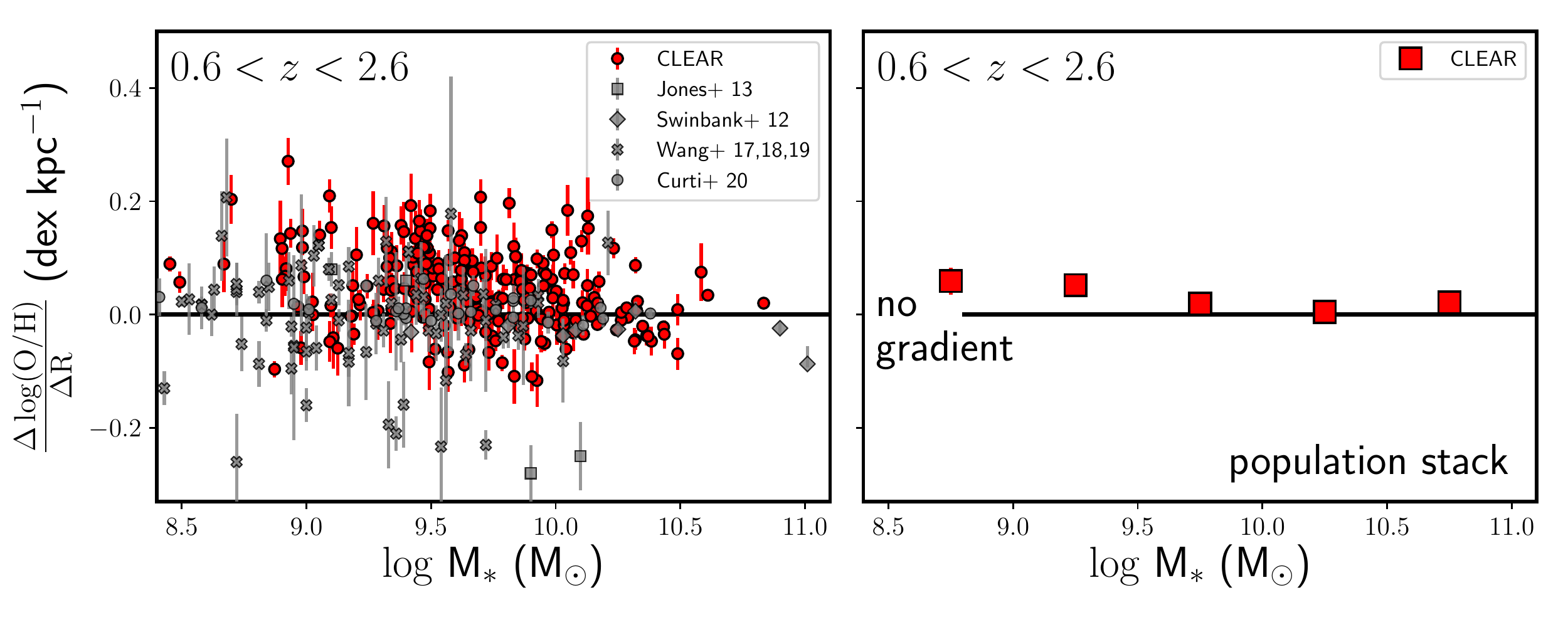}
    \caption{The metallicity gradients of the \samplesize{} galaxies in our sample are shown as a function of their stellar mass (left). The same is shown for the population-stacked metallicity maps (right), binned in stellar mass. The majority of our sample are consistent with a flat or positive metallicity gradient. A collated sample of $\sim$100 measurements from the literature are included with grey markers, taken from \citet{Swinbank12, Jones2013, 2017ApJ...837...89W, 2019ApJ...882...94W, 2020ApJ...900..183W} and \citet{Curti20}.  In the right panel, error bars show the uncertainty on the linear fits to the population-stacked profiles. In most cases, the error bars are smaller than the size of the symbols.}
    \label{fig:gradients_our_sample}
\end{figure*}

In this section, we present the main results of the paper. In \S\ref{sec:zgrad_mass} and Figures \ref{fig:gradients_our_sample} and \ref{fig:comparison_z0}, we show the gas-phase metallicity gradients of the galaxies in our sample as a function of stellar mass. We compare that with the $z\,\sim\,0$ galaxy population and highlight the inferred population evolution. In \S\ref{sec:scatter} and Figure \ref{fig:scatter}, we assess the intrinsic scatter (and its mass dependence) of the metallicity gradients. Finally, in \S\ref{sec:versus_properties} and Figures \ref{fig:dzdr_properties_combined} and \ref{fig:dzdr_correlations_significance}, we assess correlations between metallicity gradient and various galaxy properties.

\subsection{As a Function of Stellar Mass}\label{sec:zgrad_mass}

In Figure \ref{fig:gradients_our_sample}, we show the gas-phase metallicity gradients of the galaxies in our sample as a function of their stellar mass. 

In the left panel, we show individual galaxies from the CLEAR+ sample. Across all masses, the CLEAR+ galaxies are generally consistent with a flat or slightly positive gradient. Roughly 67\% of the sample (176/264 galaxies) are 3$\sigma$ consistent with a flat gradient (we define a ``flat'' gradient as objects where $\Delta\log (O/H)/\Delta R\,=0$ is within 3$\sigma$) and 29\% of the sample (77/264 galaxies) are 3$\sigma$ consistent with a positive gradient. Taken together, nearly 96$\%$ of the sample (253/264) are consistent with a flat or positive gradient and only 4\% of the sample (11/264 galaxies) are consistent with a negative gradient. With $1\sigma$ and 2$\sigma$-confidence, respectively, 84$\%$ and 89$\%$ of the sample are consistent with a flat or positive gradient.

We also include a collection of recent measurements from the literature from the following surveys at $0.8\,\lesssim\,z\,\lesssim\,2.5$: \citet{Swinbank12, Jones2013, 2017ApJ...837...89W, 2019ApJ...882...94W, 2020ApJ...900..183W} and \citet{Curti20}. The collated sample of literature measurements is shown as grey diamonds in the left panel Figure \ref{fig:gradients_our_sample}. We make no effort to correct for differences in selection and measurement technique between the samples. Our results are generally consistent with previous results in the literature: \citet{2020ApJ...900..183W} report 71$\%$ of their galaxies are consistent with a flat gradient at 2$\sigma$-confidence. \citet{Curti_2019} report 89$\%$ of their sample consistent with flat at 3$\sigma$-confidence (67$\%$ with 1$\sigma$-confidence).

In the right panel of Figure \ref{fig:gradients_our_sample}, we show the population stacks of the CLEAR sample---binned by stellar mass. The highest stellar mass bins ($9.5< \log \text{M}_{*}/\text{M}_{\odot} < 11.0$) are consistent with a flat gradient, while the lowest stellar mass bins ($8.5< \log \text{M}_{*}/\text{M}_{\odot} < 9.5$) favor a positive gradient. This result is consistent with the trends in the distribution of individual galaxies shown in the left panel.

\begin{figure}
    \centering
    \includegraphics[width =\columnwidth]{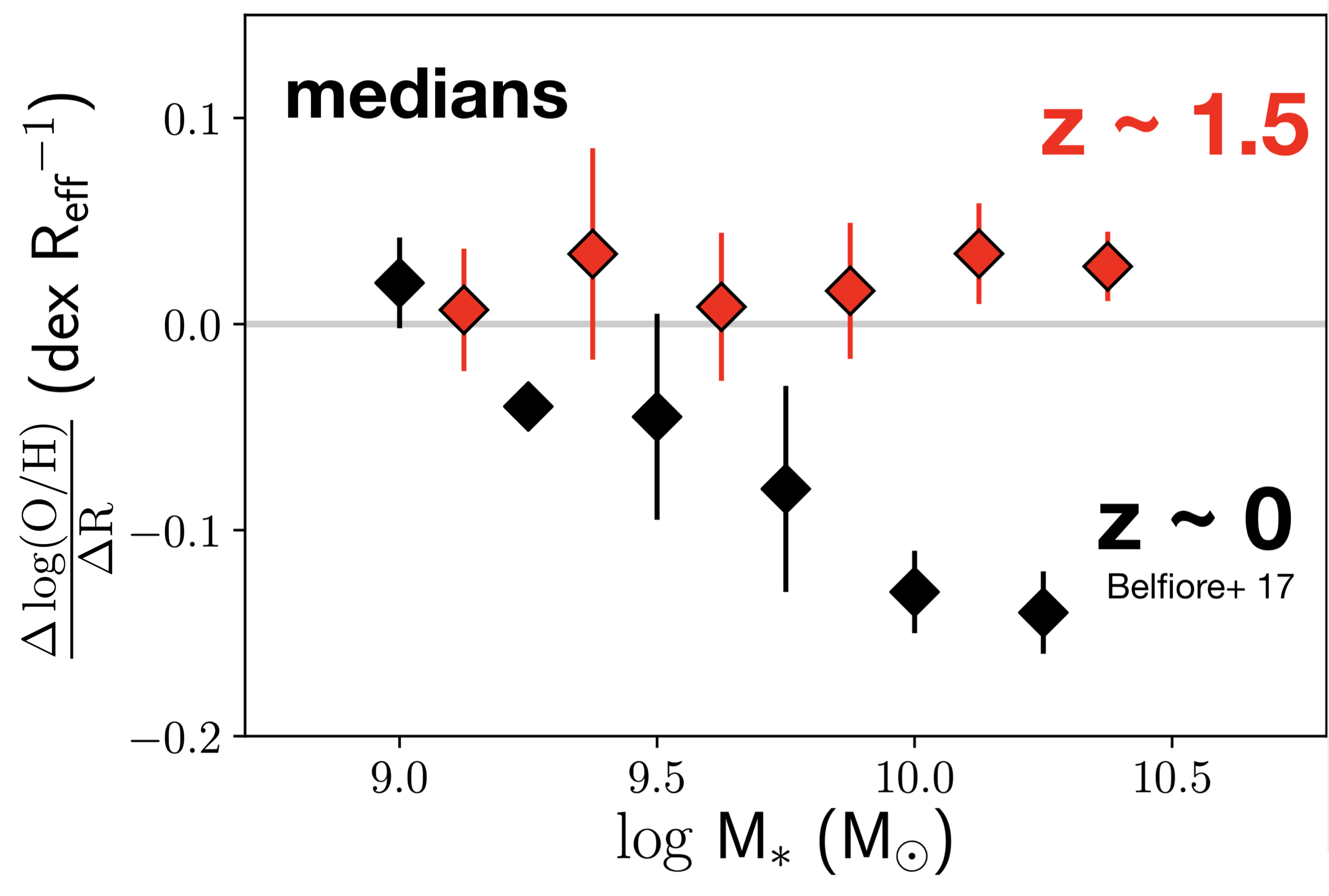}
    \caption{Population-averaged metallicity gradients versus stellar mass. Red points show the median of the CLEAR sample and the black points show a sample from \citet{2017MNRAS.469..151B} at z$\sim$0.}
    \label{fig:comparison_z0}
\end{figure}

In Figure \ref{fig:comparison_z0}, we compare our median population trends with those of star-forming galaxies at $z\sim0$ \citep {2017MNRAS.469..151B} from the MaNGA integral-field spectroscopy survey. For an apt comparison, we normalize the gradients in the CLEAR sample by their F125W (rest-frame $\sim5000$\AA) effective radius.  In the local Universe (MaNGA sample), star-forming galaxies with $\log$ M$_{*}$/M$_{\odot}$ $>$ 9.5 have radial gas-phase metallicity profiles that are declining (i.e., they have negative $\Delta \log\mathrm{(O/H})/\Delta \mathrm{R}$ gradients). The strength of the gradient decreases with decreasing mass, where the lowest mass bins ($\log$ M$_{*}$/M$_{\odot}$ $<$ 9.5) are consistent with a flat gradient. The local galaxy profiles contrast strongly with our results at $0.6 < z < 2.5$:  galaxies in this redshift range have flat, or slightly positive $\Delta \log\mathrm{(O/H})/\Delta R$ gradients across the full mass range (\massrange{}) of our sample.  At fixed mass, the differences between the $z\sim1.5$ and $z\sim0$ galaxy populations are more significant towards higher masses. At $\log$ M$_{*}$/M$_{\odot}\,\sim\,10.2$, the populations differ by $\sim$0.15 dex R$_{\text{eff}}^{-1}$.  At the low-mass end, there are no discernible differences in the metallicity gradients between the $\log$ M$_{*}$/M$_{\odot}\,\sim\,9.2$ galaxy populations.  

As an important note, the $z\,\sim\,1.5$ CLEAR+ galaxies are generally star-forming and thus still growing in stellar mass. For instance, the galaxy population with $\log$ M$_{*}$/M$_{\odot}\,\sim\,9$ at $z\,=\,1.5$ is expected to have, on average, a stellar mass of $\log$ M$_{*}$/M$_{\odot}\,\sim\,10$ by $z = 0$ \citep{2013MNRAS.428.3121M, 2017ApJ...843...46S}, and a galaxy with $\log$ M$_{*}$/M$_\odot = $10 at $z=1.5$ can double its effective radius from $2.1$~kpc to $\simeq$4 kpc by $z=0$ (e.g., \citealt{papovich15}).  As a consequence, the true evolutionary tracks of the $z\sim1.5$ galaxy populations include an increase in stellar mass coincident with a decline in metallicity gradient (i.e. towards the lower-right in Figure \ref{fig:comparison_z0})---the inferred evolution is stronger than is indicated by the population differences at fixed mass.

\subsection{Intrinsic Population Scatter}\label{sec:scatter}

\begin{figure}
    \centering
    \includegraphics[width =\columnwidth]{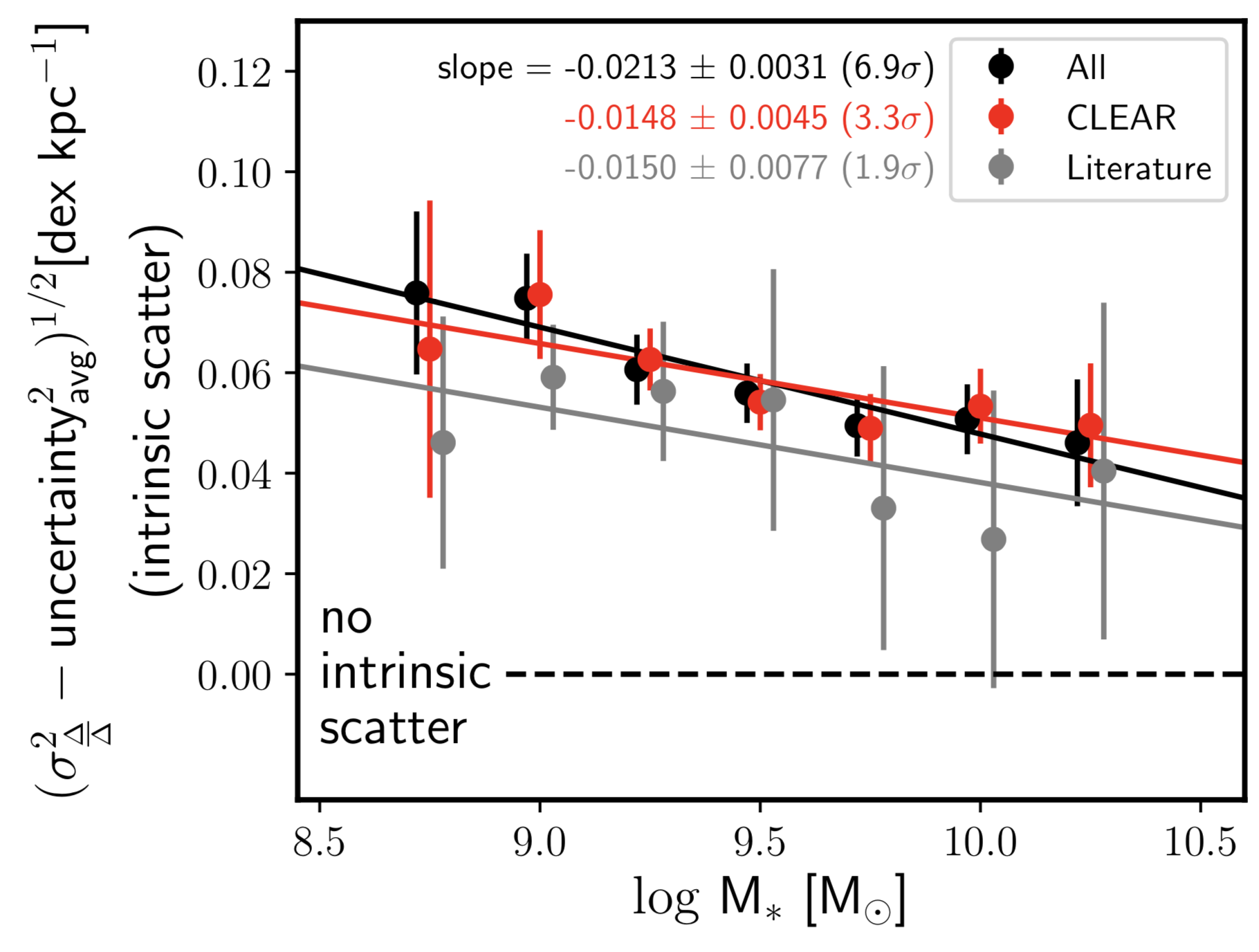}
    \caption{The inferred intrinsic scatter around the mean $\Delta\log$(O/H)/$\Delta$R relation is shown as function of stellar mass. All stellar mass bins show evidence for scatter in excess of the observational uncertainties. We report significant evidence for an increase in intrinsic scatter with a decrease in stellar mass.}
    \label{fig:scatter}
\end{figure}

In Figure \ref{fig:scatter}, we assess the intrinsic scatter of our sample and the collated literature sample as a function of stellar mass. To do so, we first measure the observed scatter using a running standard deviation of the sample. The running variable is stellar mass and we use a running width of 0.5 dex. We report the result every 0.25 dex. The conclusions below are relatively insensitive to the specific intervals chosen. To calculate the excess scatter, we measure the running median of the observational uncertainties of the sample and subtract it from the running standard deviation. We use bootstrap re-sampling to measure the standard error on the difference. At face value, this technique recovers scatter that can not be attributed to the observational uncertainties.

At all masses, we measure non-zero excess scatter, which we attribute to the intrinsic scatter of the population. The intrinsic scatter ranges from $\sim0.4-0.8$ dex kpc$^{-1}$ and it continuously increases with decreasing stellar mass. This is consistent with \citet{2020ApJ...900..183W}, which also reports an inverse correlation between stellar mass and intrinsic scatter at these redshifts.

To assess the slope of the relation between the intrinsic population scatter and stellar mass, we carry out a least-squares linear fit. We do this using: (i) the CLEAR+ sample, (ii) the collated literature sample collected in the previous subsection, and (iii) a combined CLEAR+ and literature sample. We report tentative 3.3$\,\sigma$ evidence for a slope in the CLEAR+ sample (m = -0.0148 $\pm$ 0.0045 dex kpc$^{-1}$) and 1.9$\,\sigma$ evidence in the literature sample (-0.015 $\pm$ 0.0077 dex). In the combined literature and CLEAR+ sample, the evidence for a slope increases to 6.9$\,\sigma$ (-0.0213 $\pm$ 0.003). In summary, we report excess (intrinsic) scatter in the gas-phase metallicity gradients at all masses (\massrange{}) and find strong evidence for a dependence of the intrinsic scatter on stellar mass---wherein higher mass galaxy populations have lower intrinsic scatter.

\begin{figure*}
    \centering
    \includegraphics[width=\textwidth]{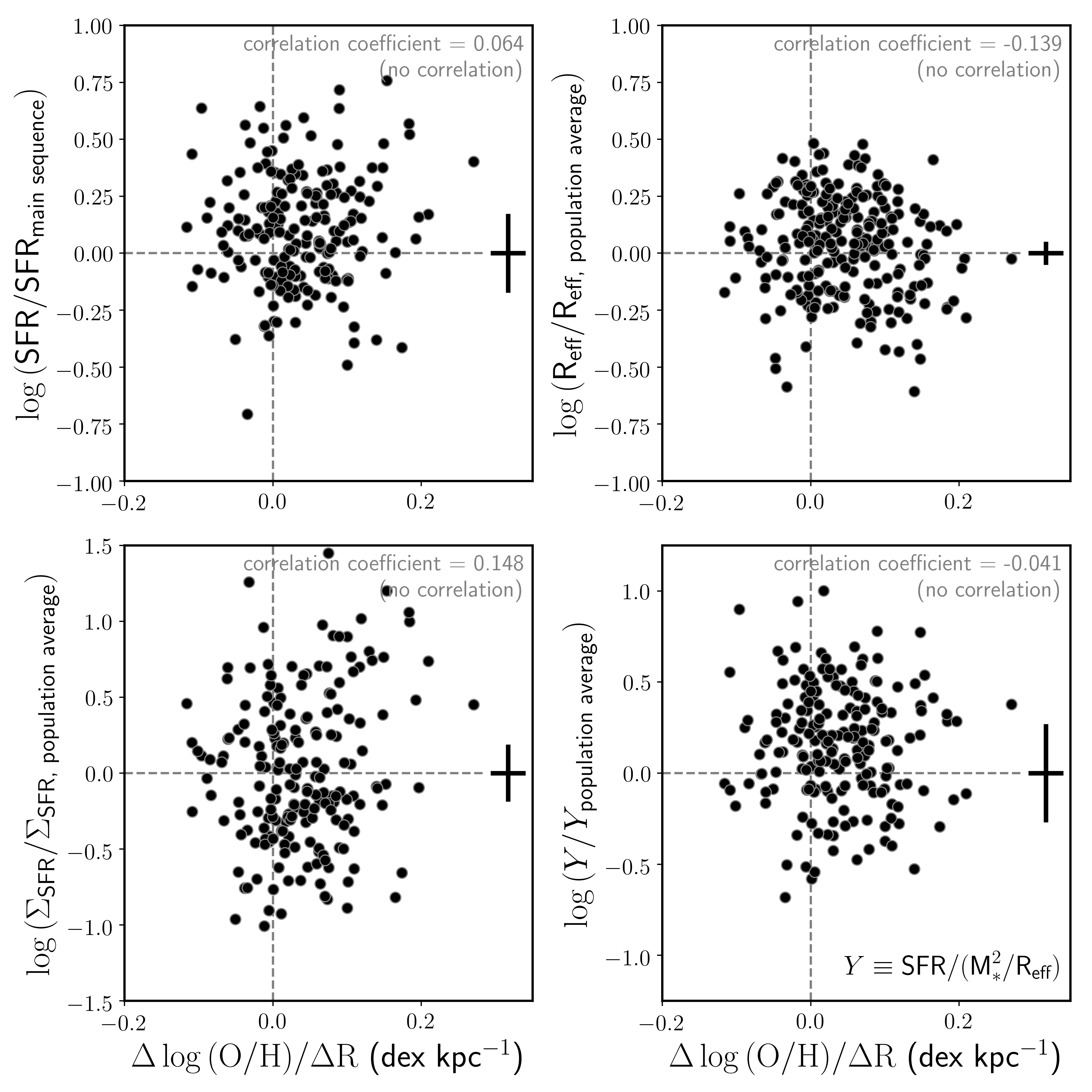}
    \caption{Correlations between gas-phase metallicity gradient and various galaxy properties:  the galaxy star-formation rate (top left), the circularized effective radius (top right), the galaxy star-formation rate surface density (bottom left), and a proxy for the star-formation rate per unit gravitational potential energy (bottom right). All of the stellar population measurements have been normalized by the population average at the mass and redshift of the galaxy. We report null correlations with all of these galaxy properties. The typical uncertainty is denoted by the errorbar shown in each panel.}
    \label{fig:dzdr_properties_combined}
\end{figure*}

\begin{figure*}
    \centering
    \includegraphics[width =\textwidth]{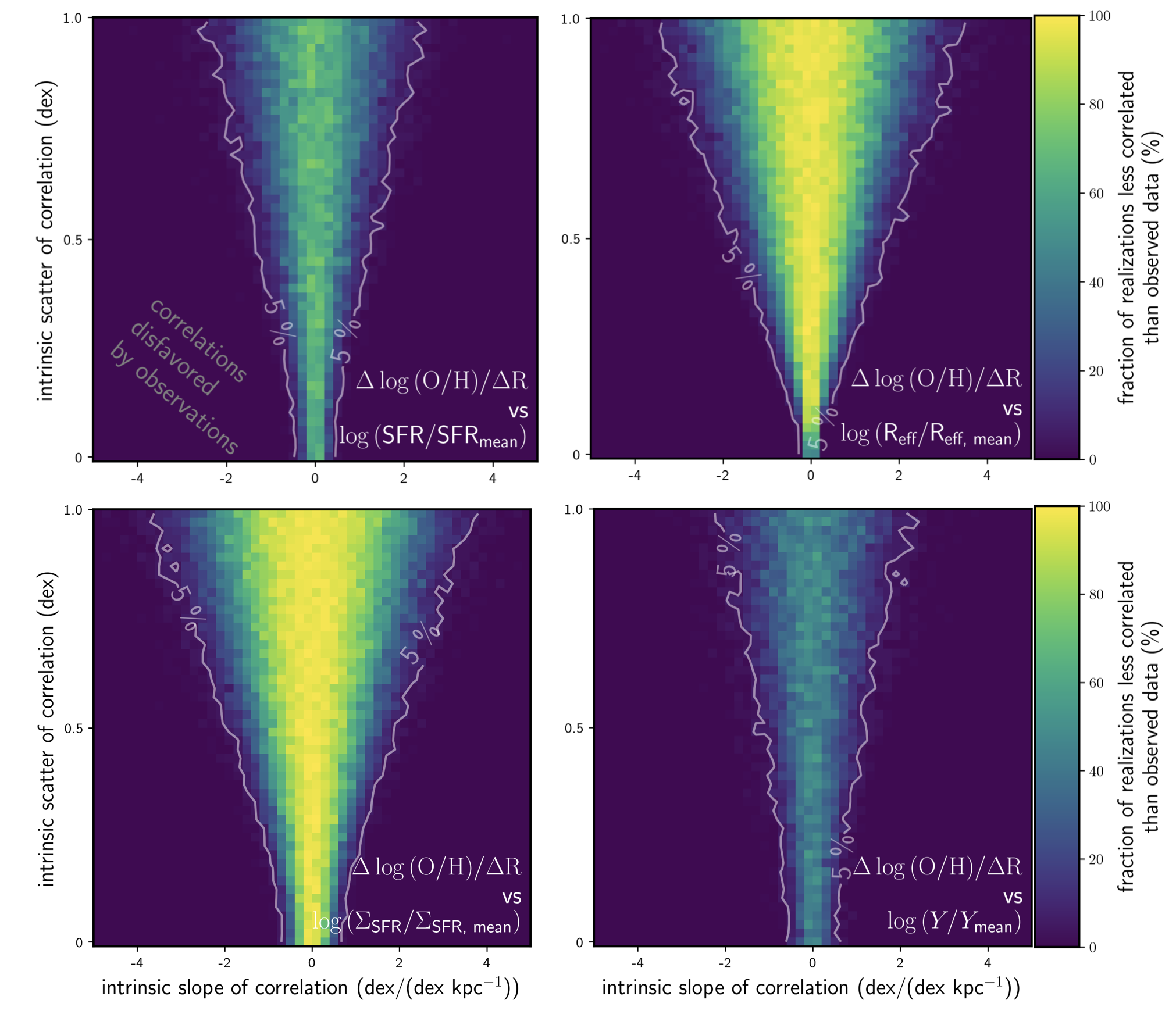}
    \caption{Correlation parameters favored---and regions of the parameter space that are ``ruled out''---by the CLEAR data.  In each region (pixel) of the plot, we simulate datasets of a fixed intrinsic correlation slope and vertical scatter. We then add noise to match the quality of the CLEAR data (as described in the text) and measure the fraction of those realizations that are \textit{less} correlated than the real data. The region outside of the contour is strongly disfavored by the CLEAR data (at 95\% confidence:  less than 5\% of the realizations fall in these regions). Specifically, this region is inconsistent with the null correlations measured between the metallicity gradients and these galaxy properties in Figure \ref{fig:dzdr_properties_combined}.}
    \label{fig:dzdr_correlations_significance}
\end{figure*}

\subsection{Correlations with Galaxy Properties}\label{sec:versus_properties}

To better understand the physical mechanisms responsible for the results in the previous two subsections, we assess correlations of the observed metallicity gradients with various galaxy properties.

In Figure \ref{fig:dzdr_properties_combined}, we use our sample to explore the relationships of metallicity gradient and several stellar population properties. Specifically, we compare the metallicity gradients of our sample to their circularized effective sizes, star-formation rates, and average star-formation surface densities (which we define as SFR/$\pi$ R$_{\text{eff}}^2$).  We also compare the metallicity gradients to the ``star-formation rate per galaxy gravitational potential energy'', which is defined as  SFR/(M$^2_{*}/$R$_{\text{eff}}$).  This latter quantity is a proxy for the ratio of available kinetic energy (from feedback associated with the SFR) to the gravitational binding energy. It allows us to test if this ratio impacts the ability of a galaxy to redistribute metals (produced from star-formation) to the rest of the galaxy, eject it into the circumgalactic medium, or unbind the gas entirely from the galaxy.  The quantities are relevant because there are observed correlations between the velocities of star-formation driven galaxy winds (usually metal-enriched) and the local star-formation surface density \citep{2012ApJ...758..135K, 2015ApJ...809..147H}. The presence (or absence) of a correlation with the last 2 parameters lends insight into the impact of star-formation winds on metallicity gradients, and/or the timescales over which they leave an imprint in the observations.

To remove the mass- and redshift-dependence of these properties, we adopt their value relative to the population-average at their mass and redshift. To do so, we use the mass-circularized size relation of late-type galaxies from \citet{vanderwel14} and the star-formation mass sequence from \citet{2014ApJ...795..104W}. 

We measure the Pearson coefficient for each correlation and report the value (and an interpretation) in the top right of each panel. We report no evidence for a correlation between metallicity gradient and any of these mass-normalized physical properties. 

As an aside, we do note that the galaxies with the highest star-formation rate surface densities (those that are a dex above the population average) tend to have more positive metallicity gradients than the rest of the sample (bottom left panel, Figure \ref{fig:dzdr_properties_combined}). A plausible physical explanation that can account for this result {\emph{and}} the measured null correlation is that the star-formation rate surface densities need to reach a certain threshold before they are able to effectively launch metal-rich winds (see e.g., \citealt{2015ApJ...809..147H}). If such winds are preferentially launched from the galaxy center, then they could presumably drop the central metallicity and push the metallicity gradient in the positive direction.

Figure~\ref{fig:dzdr_properties_combined} shows the results for the galaxies in our sample. Again, we find no statistically-significant evidence for a correlation between the metallicity gradients and these galaxy properties.  However, even these null/weak correlations place important constraints on predictions from theoretical models.  We discuss this in more detail below. 

In Figure \ref{fig:dzdr_correlations_significance}, we estimate the intrinsic correlations that are {\emph{ruled out}} by the CLEAR observations---specifically, those that are inconsistent with the null observed correlations between the metallicity gradient and each of the galaxy properties above. 

To do so, we simulate ``CLEAR-like" realizations of a galaxy population assuming there is some intrinsic correlation between the metallicity gradient and each galaxy parameter. The correlations are defined by two parameters: the intrinsic slope of the correlation, and the intrinsic vertical scatter of the correlated variables. For each simulated intrinsic correlation (i.e., each pixel in the Figure \ref{fig:dzdr_correlations_significance} heatmaps), we generate 100 sample realizations which are then added noise to match the quality of the CLEAR+ measurements. Specifically, we draw a mock sample using the intrinsic correlation parameters with the same dynamic range of metallicity gradients as the CLEAR+ sample, and the same number of galaxies. We add simple uncorrelated Gaussian noise to mimic the observational uncertainties. Finally, we measure the Pearson correlation coefficient for each realization.

In Figure \ref{fig:dzdr_correlations_significance}, we show the fraction of simulated realizations that are as (or more) uncorrelated than the observed CLEAR sample. To interpret these diagrams, the parameter space where the fractions are high are consistent with the CLEAR observations---e.g., a fraction of 80$\%$ means that 80$\%$ of the realizations are as uncorrelated as the observed correlations. The parameter space where the fractions are low are strongly disfavored by the CLEAR results. The region of the diagram outside the 5$\%$ contour shown in the diagrams are ruled out at $2\sigma$ confidence or more. The results shown in Figure \ref{fig:dzdr_correlations_significance} offer a direct constraint on theoretical predictions as our data rule out these regions of the parameter space at 95\% confidence.

\begin{figure}
    \centering
    \includegraphics[width = \columnwidth]{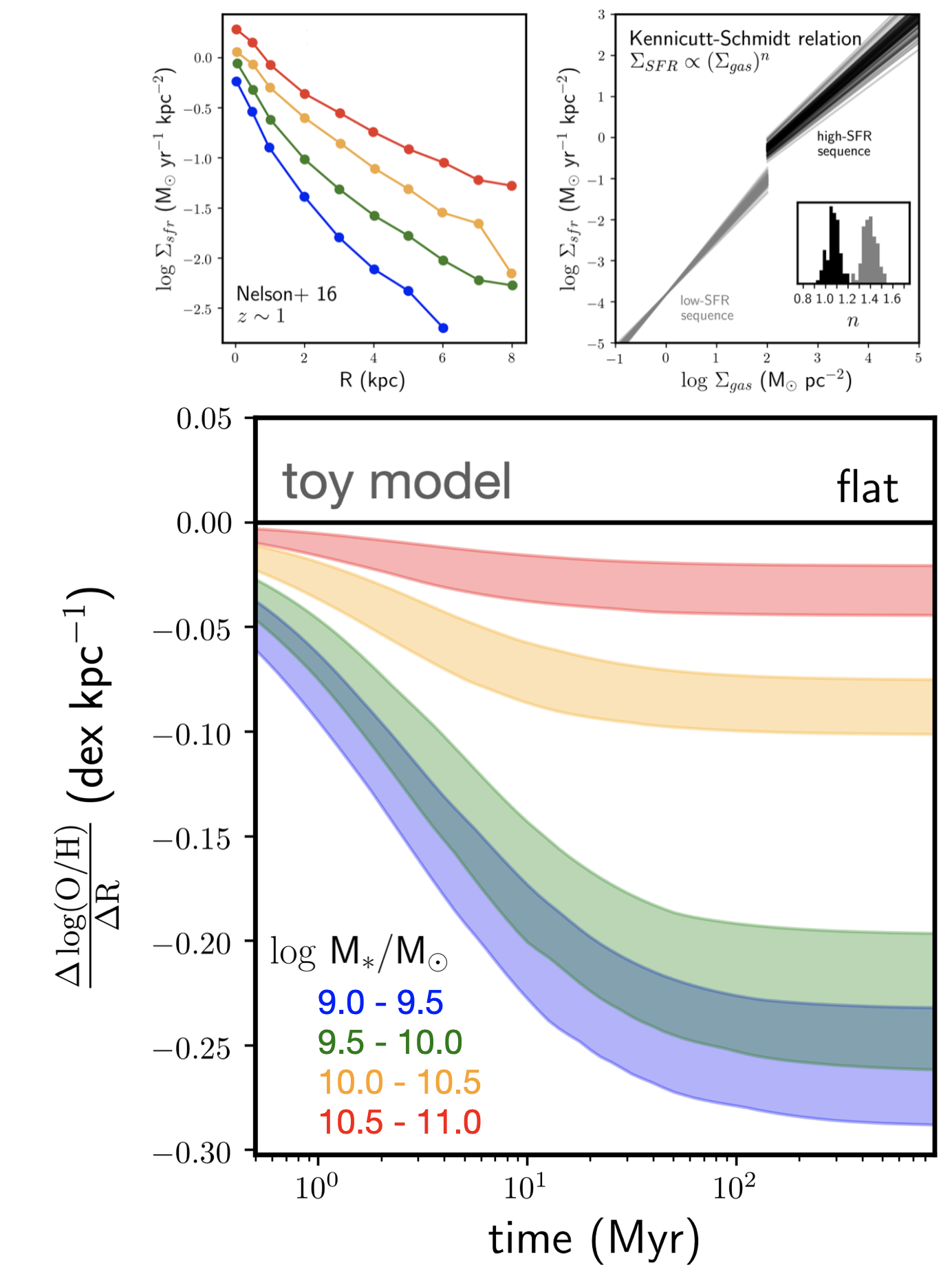}
    \caption{A simple prediction for the time evolution of metallicity gradients ($\Delta \log \mathrm{(O/H)}/\Delta \mathrm{R}$) in star-forming galaxies at $z\sim1$ (bottom panel).  The toy model is governed by the observed radial profiles of the star-formation rate density in $z\sim1$ galaxies (\citealt{2016ApJ...828...27N}; top left) and a broken-power law Kennicutt-Schmidt relation (\citealt{1998ApJ...498..541K}; top right). The toy model starts the simulation at $t=0$ with a flat gradient and a metallicity matching the $z\sim2.3$ mass-metallicity relation \citep{2018ApJ...858...99S}. The shaded regions show the results from this toy-model. Without additional metal redistribution, we should {\emph{expect}} galaxies to rapidly develop declining metallicity gradients through star-formation and stellar evolution---and to do so faster in lower mass galaxies. The ubiquity of flat gradients in the observations implies that gradients are being disrupted on short timescales ($\sim10-100$ Myr). The shading of each line represents the scatter of 100 realizations with varying KS slopes and metal yields (see text for details).}
    \label{fig:toy_model}
\end{figure}

\section{Discussion}\label{sec:discussion}

Metals are a dye for galaxies. They are formed in stars, deposited around young star-forming regions, and swept up in SNe/massive-star-driven winds. As gas moves around galaxies, so too will its entrained metals (or, equivalently, its lack of metals). If the spatial distribution of the gas phase metallicity strays from the distribution of a galaxy's stars (an integrated record of past star-formation and metal enrichment), it indicates one (or both) of the following: (i) the metals that formed in the galaxy were re-distributed from their birthplace or (ii) the interstellar medium (ISM) of the galaxy was un-evenly diluted through metal-poor gas accretion. In either case, the evolution of the metallicity gradient is intimately linked with galaxy- and halo-scale gas flows.

In this section, we develop and test a toy model to highlight the rapid evolutionary timescales implied by the observed metallicity gradients (\S\ref{sec:toy_model} and Figure \ref{fig:toy_model}) and discuss various processes known to flatten metallicity gradients (\S\ref{sec:flattening_gradients}).

\subsection{A Toy Model: Timing the Development of Metallicity Gradients at $z\sim1$}\label{sec:toy_model}

That a large fraction of galaxies have flat and inverted metallicity gradients at \redshiftrange{} (\S\ref{sec:zgrad_mass}) strongly favors a scenario in which metallicity gradients are destroyed in galaxies on short timescales---quicker than the short time it takes for star-formation to develop an (observable) declining gradient. In Figure \ref{fig:toy_model}, we develop an empirical toy model to illustrate this. 

\noindent{\bf{Constructing the Model.}} We adopt a set of empirical prescriptions to evolve the radial metallicity profiles of mock galaxy populations. For simplicity, these models assume metal production from star-formation, no radial mixing of metals, and an instantaneous replenishment of the gas used to form stars. The empirical prescriptions include (i) the average radial profiles of star-formation rate surface density at $z\sim1$ measured from the 3D-HST survey \citep{2016ApJ...828...27N}, (ii) the global mass-metallicity relation at $z\,\sim\,2.3$ \citep{2018ApJ...858...99S}, and (iii) the Kennicutt-Schmidt (KS) law \citep{1998ApJ...498..541K} relating the star-formation and gas mass surface densities. The first dictates the rate at which stars form per galactic annulus and, thus, the rate at which metals are deposited into the local interstellar medium. The second dictates the global normalization of the metallicity at the start of the simulation. The third dictates the differential increase in metallicity per unit of star-formation. If the KS relation has a power-law slope that exceeds 1, it means that stars are forming more efficiently at higher gas densities. As a result, star-formation would more efficiently increase the local ISM metallicity at higher gas densities. This dependence of the rate of change of {\emph{metallicity}} (a subtle difference from the absolute metal production rate) on gas density is key to the model. 

The KS law ($\Sigma_{\text{SFR}}\,\propto\,\Sigma_{\text{gas}}^n$) at high surface densities at these redshifts has a slope between unity and 2 \citep{2011MNRAS.412..287N, 2013ApJ...768...74T}. We adopt a probabilistic broken power-law for the KS relation in our model, reflecting a regular and starburst sequence. The break occurs at 100 M$_{\odot}$ pc$^{-2}$, above which the relation assumes a flatter slope. The slope above and below the power law break are drawn from a Gaussian distribution with a mean of $n=1.1$ and $n=1.4$, respectively, and a width of 0.1. We also adopt a probabilistic metal yield from type II SNe, using a Gaussian with a mean of y$_{z, ii}$ = 0.03 and a width of 0.005. We simulate 4 galaxy populations of $\log$ M$_*$/M$_{\odot}$ = [$9.0-9.5$, $9.5-10.0$, $10.0-10.5$, $10.5-11.0$]. Lastly, we assume that the ratio of oxygen to total metals is constant, such that the change in the oxygen abundance gradient is equivalent to the change in the metallicity gradient. For each population, we simulate 100 realizations---each using a random draw from the distributions of KS slope and metal yield. We evolve each simulation for 1 Gyr. The initial metallicity gradient is set as flat with a global metallicity matching the $z\,\sim\,2.3$ mass metallicity relation \citep{2018ApJ...858...99S}.

This is a simple empirical model, and should not be confused with more sophisticated chemical evolution models (e.g., \citealt{2011A&A...531A..72S, 2019MNRAS.490..665M,  2020MNRAS.491.5795H}). With that said, it illustrates an important point. 

\noindent{\bf{Toy Model Predictions.}} Given the observed radial profile of star-formation and the KS relation at $z\sim1$, Figure \ref{fig:toy_model} indicates that the {\emph{rate of change of metallicity}} due to star-formation should be higher in the centers of galaxies than in their outskirts. 

Furthermore, if metals were stationary (in a radial sense), we anticipate galaxies should develop {\emph{detectable}} (i.e., $\Delta \log \mathrm{(O/H)}/\Delta \mathrm{R}$ $<\,-0.05$) metallicity gradients rapidly---on the order of $10-100$ Myr in galaxy sub-populations below $\log$ M$_*$/M$_{\odot}\,<\,10.5$. To reiterate, the fact that we observe a large fraction of flat and positive gradients indicates that declining metallicity gradients must be destroyed in galaxies on timescales shorter than the short time it takes for them to establish.

In Figure \ref{fig:toy_model}, the higher mass galaxy populations exhibit shallower evolution than the lower mass galaxy populations. This is an (informative) consequence of the broken power law form of the Kennicutt-Schmidt relation we adopt. The lower mass galaxies tend to draw more from the low $\Sigma_{\text{gas}}$ branch of the KS relation (i.e., the branch with the steeper $n\sim1.4$ slope), while the higher mass galaxies tend to draw more from the high $\Sigma_{\text{gas}}$ branch (i.e., the branch with the shallower $n\sim1.1$ slope). The shallower the slope in the KS relation, the less of a gradient develops. At $n = 1$, the rate of change of the metallicity is the same everywhere---no gradient would develop.

\subsection{Flattening Metallicity Gradients}\label{sec:flattening_gradients}

In Figure \ref{fig:comparison_z0}, we infer a strong mass-dependent evolution in metallicity gradients from $z\sim1.5$ to today---the evolution is steeper in higher mass galaxies.

This result has an enticing parallel with the observed kinematic evolution of star-forming galaxies from $z\sim2$ to now. At $z\sim2$, the velocity dispersions of the ionized gas in galaxies are several factors higher than they are today \citep{2007ApJ...660L..35K, 2015ApJ...799..209W, 2016ApJ...830...14S, 2019ApJ...880...48U}. High gas velocity dispersions are likely accompanied by efficient radial mixing---acting to re-distribute metal-enriched gas from the centers of galaxies to their outskirts. With time, star-forming galaxies gradually increase in rotational support and decline in dispersion support \citep{2012ApJ...758..106K, 2017ApJ...843...46S}. This evolution is a strong function of mass---with less massive galaxy populations on average having higher contributions from dispersion support at all times.

These kinematic parallels extend down to $z\,\sim\,0$. In the local universe, the majority of massive star-forming galaxies ($\log \text{M}_*/\text{M}_{\odot}\,>\,9.5$) have formed a rotationally-supported disk, whereas only a fraction of low mass star-forming galaxies ($\log \text{M}_*/\text{M}_{\odot}\,<\,9.5$) have formed a disk \citep{2015MNRAS.452..986S}. This lingering dispersion support in the ionized gas in low mass galaxies today may contribute to the average flatness of their gas-phase metallicity profiles \citep{2019MNRAS.487..456B}.

Using the FIRE galaxy formation simulations, \citet{2017MNRAS.466.4780M} conclude that strong (declining) metallicity gradients only appear in galaxies with a well-formed disk (i.e., a galaxy with a rotation velocity higher than its local velocity dispersion)---but not all well-formed disks have a strong metallicity gradient. Furthermore, they find that highly perturbed  non-rotating (mostly post-merger)  galaxies tend to have flat gradients. 

In \S\ref{sec:scatter}, we report evidence for an increase in the intrinsic scatter of the metallicity gradients towards lower masses. This is consistent with recent results at this redshift \citep{2020ApJ...900..183W}, and with observations of the local universe \citep{2015MNRAS.448.2030H, 2019MNRAS.488.3826B}. The scatter in metallicity gradients provides a key benchmark for galaxy evolution models.

Specifically, the mean and scatter of the metallicity gradients has been shown to be a sensitive probe of stellar feedback. Using closed-box chemical evolution models, \citet{2015MNRAS.448.2030H} demonstrate that the population mean and scatter of metallicity gradients is sensitive to mass accretion rates and mass-loading factors---with high rates of both producing a narrow distribution centered around flat gradients (see also \citealt{2019MNRAS.487..456B}).

\citet{2013A&A...554A..47G} compared the metallicity gradients of simulated galaxies from two simulation suites---one run with a `conservative' feedback model and one with an `enhanced' feedback model that can more efficiently drive hot SNe winds. Galaxies forming in the simulation with enhanced feedback always have flat metallicity gradients, while galaxies forming in the simulation with conservative feedback always rapidly build up metallicity gradients at high redshift and flatten at late times as the galaxy grows. On the other hand, in the FIRE simulations, \citet{2017MNRAS.466.4780M} recover a wide scatter in metallicity gradients all with the same feedback model. They attribute this, in part, to the burstiness of star-formation in the FIRE galaxies, and the ability of the sub-kpc feedback model to switch between favorable and unfavorable conditions for driving metal-enriched outflows. \citet{2020arXiv200710993H} study the gas-phase metallicity gradients in the Illustris-TNG simulations, and find that the Illustris-TNG galaxies generally have steeper gradients than those found in the FIRE simulations. These differences are attributed, at least in part, to the differences in the feedback models---the Illustris-TNG model leads to less bursty and disruptive behavior than the FIRE feedback model.

It is clear that statistical measures of metallicity gradients at high redshift serve as an important benchmark for galaxy formation feedback models. Gradient demographics provide a unique way to confront the physical models underpinning these numerical simulations.

Another important actor for setting and disrupting metallicity gradients is gas accretion---either through metal-poor accretion from intergalactic filaments, (re-)accreted material from the circumgalactic medium, or galaxy mergers. Each channel acts in a unique manner. The impact of metal-poor accretion from cold filaments strongly depends on where that material is deposited. If deposited directly into the centers of galaxies (or radially-mixed on quick timescales), it should act to dilute the central metallicity and flatten metallicity gradients. If the metal-poor gas is deposited on the outskirts of galaxies, it should push the gradient negative. For (re-)accreted material, a general flattening of metallicity gradients will occur as metals ejected into the CGM rain back onto the outskirts of galaxies \citep{2013MNRAS.434.1531F, 2019MNRAS.490.4786G}. For mergers, numerical simulations indicate that they can rapidly re-distribute metals around galaxies and flatten metallicity gradients \citep{2010ApJ...710L.156R, 2011MNRAS.417..580P, 2012ApJ...746..108T}. In general, accretion and merger rates are expected to be factors of $10-30$ higher at $z\sim2$ than they are today \citep{2015MNRAS.449...49R, 2017ApJ...837..150S}. 

As the prevalence of all of these `flattening mechanisms' (star-formation, accretion, mergers) declines with time, it is reasonable to assume that there will be a corresponding decline in the rate in which metals are (re-)distributed around galaxies and the interstellar medium is diluted by metal-poor accretion. At late times, in a more hospitable universe, galaxies will have the chance to develop long-lived {\emph{declining}} metallicity gradients through star-formation.

\section{Conclusions}\label{sec:conclusions}
We report on the gas-phase metallicity gradients in \samplesize{} galaxies over \redshiftrange{}.  The observations include deep near-infrared {\emph{HST}}/WFC3-G102 grism spectra taken through the CANDELS Ly$\alpha$ Emission at Reionization (CLEAR) survey, as well as publicly-available WFC3-G102+G141 grism spectra overlapping the CLEAR footprint. The combined G102 + G141 spectral coverage allows for simultaneous coverage of multiple strong-line metallicity indicators (notably [O\ii], [O\iii], and H$\beta$); and the high spatial resolution of the grism enables for resolved maps at the resolution of {\emph{HST}}. 

We summarize our conclusions as follows:

\begin{itemize}
\item The majority of the galaxies (84$\%$) in our sample have flat or positive radial gas-phase metallicity gradients. This result is generally consistent with other results in the literature (e.g., \citealt{2020ApJ...900..183W, Curti20}), and is in marked contrast with the $z\sim0$ galaxy population \citep{2017MNRAS.469..151B}. We produce an empirical toy model to demonstrate that flat and positive metallicity gradients require gas phase metals to be (re-)distributed, or central metallicities to be diluted from metal-poor gas accretion, on rapid $\sim$10-100 Myr timescales.
    
\item We detect intrinsic population scatter---i.e., scatter in excess of that which is expected from the observational uncertainties---across our full mass range (\massrange{}). We find strong evidence for a mass dependence to the scatter (such that there is excess scatter at lower masses) in the CLEAR sample (3.3$\sigma$) and even stronger evidence when we combine with existing measurements in the literature (6.9$\sigma$).
    
\item We explore correlations between the gas-phase metallicity gradient and stellar population properties at fixed stellar mass, including: star-formation, sizes, star-formation surface density, and star-formation per potential energy. We find no statistically-significant correlations between the metallicity gradient and any of these properties. We place constraints on the parameter space of intrinsic correlations ruled out by the CLEAR data. These constraints can be readily confronted against predictions from theoretical models.
    
\end{itemize}

These results strongly favor a scenario in which metals are re-distributed around galaxies on faster timescales than the short time it should take for star-formation and stellar evolution to lead to a declining metallicity gradient. Despite rapid progress on both the observational and theoretical front, a full description of the processes responsible for this (re-)distribution---specifically, their timescales and relevance as a function of galaxy mass and redshift---remains outstanding.

\section*{Acknowledgements}

We thank our colleagues on the CLEAR team for their
valuable conversations and contributions. RCS thanks Xin Wang, Ayan Acharyya, Erini Lambrides, Alaina Henry, Susan Kassin, and Alexander de la Vega for valuable conversations.
RCS appreciates support from a Giacconi Fellowship at the Space Telescope Science Institute. VEC acknowledges support from the NASA Headquarters under the Future
Investigators in NASA Earth and Space Science and Technology (FINESST)
award 19-ASTRO19-0122, as well as support from the Hagler Institute
for Advanced Study at Texas A\&M University.  CP, VEC, and JM acknowledge generous support from the George P. and Cynthia Woods Mitchell Institute for Fundamental Physics and Astronomy. This work is based on data obtained from the Hubble Space Telescope through program number GO-14227. Support for Program number GO-14227 was provided by NASA through a grant from the Space Telescope Science Institute, which is operated by the Association of Universities for Research in Astronomy, Incorporated, under NASA contract NAS5-26555. This research made use of Astropy,\footnote{http://www.astropy.org} a community-developed core Python package for Astronomy \citep{astropy:2013, astropy:2018}.

\bibliography{CLEAR_gradients.bib}{}

\end{document}